\documentclass[10pt,letterpaper]{article}
\usepackage[top=0.85in,left=2.75in,footskip=0.75in]{geometry}

\usepackage{changepage}

\usepackage[utf8]{inputenc}

\usepackage{textcomp,marvosym}

\usepackage{fixltx2e}

\usepackage{amsmath,amssymb}

\usepackage{cite}

\usepackage{nameref,hyperref}

\usepackage[right]{lineno}

\usepackage{microtype}
\DisableLigatures[f]{encoding = *, family = * }

\usepackage{rotating}

\raggedright
\usepackage[aboveskip=1pt,labelfont=bf,labelsep=period,justification=raggedright,singlelinecheck=off]{caption}

\makeatletter
\renewcommand{\@biblabel}[1]{\quad#1.}
\makeatother

\date{}

\usepackage{lastpage,fancyhdr,graphicx}
\usepackage{epstopdf}
\fancyheadoffset[L]{2.25in}
\fancyfootoffset[L]{2.25in}
\newcommand{\filename}[1]{\textsf{\textbf{#1}}}

\begin{document}
\vspace*{0.35in}

% Please capitalize all terms in the title except conjunctions, prepositions, and articles.
\begin{flushleft}
{\Large
\textbf\newline{Target-distractor Synchrony Affects Performance in a
  Novel Motor Task for Studying Action Selection}
}
\newline
\\
Sebastian James\textsuperscript{1,2},
Olivia A. Bell\textsuperscript{1,\Yinyang},
Muhammed A. M. Nazli\textsuperscript{1,\Yinyang},
Rachel E. Pearce\textsuperscript{1,\Yinyang},
Jonathan Spencer\textsuperscript{1,\Yinyang},
Katie Tyrrell\textsuperscript{1,\Yinyang},
Phillip J. Paine\textsuperscript{5},
Timothy J. Heaton\textsuperscript{5},
Sean Anderson\textsuperscript{3,2},
Mauro Da Lio\textsuperscript{4,\ddag},
Kevin Gurney\textsuperscript{1,2,*,\ddag}
\\
\bigskip
\bf{1} Adaptive Behaviour Research Group, Department of Psychology, The University of Sheffield, Sheffield, United Kingdom
\\
\bf{2} Insigneo Institute for in-silico Medicine, The University of Sheffield, Sheffield, United Kingdom
\\
\bf{3} Department of Automatic Control Systems Engineering, The University of Sheffield, Sheffield, United Kingdom
\\
\bf{4} Department of Industrial Engineering, Universit\`{a} degli Studi di Trento, Trento, Italy
\\
\bf{5} School of Mathematics and Statistics, The University of Sheffield, Sheffield, United Kingdom
\\\bigskip

% Insert additional author notes using the symbols described below. Insert symbol callouts after author names as necessary.
%
% Remove or comment out the author notes below if they aren't used.
%
% Primary Equal Contribution Note
\Yinyang These authors contributed equally to this work.

% Additional Equal Contribution Note
% Also use this double-dagger symbol for special authorship notes, such as senior authorship.
\ddag These authors contributed equally to this work.

% \textcurrency a Insert current address of first author with an address update
% \textcurrency b Insert current address of second author with an address update

% Group/Consortium Author Note
% \textpilcrow Membership list can be found in the Acknowledgments section.

% Use the asterisk to denote corresponding authorship and provide email address in note below.
* k.gurney@sheffield.ac.uk

\end{flushleft}
\section*{Abstract}
The study of action selection in humans can present challenges of task
design since our actions are usually defined by many degrees of
freedom and therefore occupy a large action-space. While saccadic
eye-movement offers a more constrained paradigm for investigating
action selection, the study of reach-and-grasp in upper limbs has
often been defined by more complex scenarios, not easily interpretable
in terms of such selection.  Here we present a novel motor behaviour
task which addresses this by limiting the action space to a single
degree of freedom in which subjects have to track (using a stylus) a
vertical coloured target line displayed on a tablet computer, whilst
ignoring a similarly oriented distractor line in a different
colour. We ran this task with 55 subjects and showed that, in
agreement with previous studies, the presence of the distractor
generally increases the movement latency and directional error
rate. Further, we used two distractor conditions according to whether
the distractor's location changes asynchronously or synchronously with
the location of the target. We found that the asynchronous distractor
yielded poorer performance than its synchronous counterpart, with
significantly higher movement latencies and higher error rates. We
interpret these results in an action selection framework with two
actions (move left or right) and competing `action requests' offered
by the target and distractor. As such, the results provide insights
into action selection performance in humans and supply data for
directly constraining future computational models therein.

%\linenumbers

\section*{Introduction}

% Describes the idea that organisms must select between
% actions. Targets and distractors.
Selecting an appropriate target for action in an unpredictable
environment is crucial to survival for all
animals~\cite{cisek_cortical_2007,pratt_action-centered_1994}. As an
example, consider the choice between grasping for a nutritious blue
fruit whilst ignoring the distraction of poisonous red berries. Such
an interaction is dependent upon an integration of perception and
action~\cite{tipper_actionbased_1998}. The simple task of perceiving
the target berry and grasping it is typically coordinated by a rapid
eye movement toward the target, followed by goal-directed hand and arm
movements~\cite{biguer_coordination_1982,neggers_ocular_2000,neggers_gaze_2001}.
Individuals must have the ability to rapidly distinguish between
relevant and irrelevant information within the environment, and
selectively initiate the most suitable movement; a process referred to
as \emph{action selection}; a term which began to be used within the
context of biological motor control in the late
1980s~\cite{norman_attention_1986,maes_dynamics_1989,maes_situated_1990}.

The
ability to select the appropriate movement requires the individual to
choose between an abundance of conflicts within competing brain
systems~\cite{redgrave_basal_1999}. In our example, the blue berries
must be selected as the grasp target in preference to the red berries,
but it may also be necessary for the animal to select the action of
fleeing if it detects the approach of a predator or returning to the
safety of its nest if it is sated or tired. Action selection may refer
either to the within-goal selection of a movement (towards the blue
berry or towards the red berry) or to a selection between different
goals (continue feeding or return to nest).

% paragraph removed from here and replaced with:
In this paper, we are concerned with the former, within-goal selection
choice. We wanted to investigate motor control and action selection,
and to this end we developed a task in which we could measure reaction
times, error rates and trajectories of stylus movements towards a
target line in the presence of a differently coloured distractor
line. Here, we present the task, which we call the \emph{line task},
along with a sample implementation of the task with a healthy cohort
of subjects.

For many years, the saccadic system has been used for the
investigation of action selection and decision
making~\cite{deubel_saccade_1996,mcpeek_saccade_2002,carpenter_contrast_2004,rorie_integration_2010,bompas_saccadic_2011,reppert_modulation_2015}. A
well structured set of experimental protocols has been developed by
researchers in this field. Pro-saccades, for which the subject must
move their eyes towards a target, are elicited in `gap, step and
overlap' paradigms~\cite{saslow_effects_1967}. The anti-saccade, in
which the subject is required to saccade \emph{away} from a target,
has a proposed
protocol~\cite{antoniades_internationally_2013}. Saccades provide a
good window through which to view sensorimotor decisions because they
can be considered to have only two degrees of freedom and the anatomy
is reasonably well
understood~\cite{moschovakis_anatomy_1994,moschovakis_microscopic_1996,sparks_brainstem_2002}.
The predictive properties of the superior
colliculus~\cite{ratcliff_comparison_2003}, frontal eye
field~\cite{schall_neural_1993} and lateral intraparietal
area~\cite{hanks_microstimulation_2006,rorie_integration_2010} have
all been investigated.  However, the saccadic system is highly
specialised for eye movements and an understanding of it may not
generalise to motor control of limb movements. Furthermore, saccades
and reaching decisions are closely
connected~\cite{gribble_hand-eye_2002}. Although an observation of a
primate making a reaching movement will show that the saccade precedes
the reach (as described in the berries example), internal neural
control processes for the two movements are not necessarily
sequential~\cite{carey_eyehand_2000} and the reaching movement has
been shown to be specified prior to the onset of the
saccade~\cite{gribble_hand-eye_2002}. Reaching movements affect both
saccade trajectories~\cite{tipper_reaching_2001} and
latencies~\cite{fisk_organization_1985,neggers_ocular_2000,snyder_eye-hand_2002}
and they may be planned in an eye-centered coordinate
system~\cite{batista_reach_1999,tipper_reaching_2001}. Because of this
close relationship, it is important to conduct behavioural studies
using both reach and saccadic eye movements to reveal internal,
decision making processes.

The simplicity of saccadic paradigms is usually lost in experiments on
limb control. Thus, many reach to grasp experiments involve the
recognition of targets and distractors within a three dimensional
space. Consequently, higher-order visual processes such as depth
perception are employed~\cite{welsh_movement_2004}. Additionally,
tasks requiring prehension movements involve the recognition of object
size, orientation and shape~\cite{meegan_visual_1999}. Both the
ventral and dorsal streams are therefore likely to be engaged whilst
completing reach and grasp tasks~\cite{milner_two_2008}. As well as
high order visual processing, these tasks require intricate grasp
movements and complex motion throughout a three dimensional
space~\cite{howard_hand_1997,tipper_selective_1997,
  tipper_actionbased_1998,castiello_mechanisms_1999,jackson_are_1995}. A
consequence of the complexity associated with these tasks is that
participants may be selecting from not a few, but from many competing
actions or affordances~\cite{cisek_cortical_2007}.

% New section on step tracking experiments.
There does exist a class of experiments known as \emph{step-tracking
  tasks} which study reach movements in response to simple, quick
movements of a target. For example, in one early, electronic
trajectory-tracking apparatus, Trumbo and colleagues displayed a
vertical hairline target on the 5 inch screen of an
oscilloscope~\cite{trumbo_versatile_1963}. Subjects were required to
match the location of a second line to the target by controlling the
position of their arm which was fixed but allowed to pivot at the
elbow~\cite{trumbo_motor_1968}. Hallett and co-workers used a similar
apparatus to that used by Trumbo et al.~in an investigation of EMG
activity in biceps and triceps muscles~\cite{hallett_emg_1975}. Other
researchers performed step-tracking experiments for wrist movements,
tracking a target moving in one
dimension~\cite{waters_influence_1981,hoffman_step-tracking_1986} or
two~\cite{haruno_optimal_2005}. In each of these step-tracking
experiments the subject would move a manipulandum lever to position a
cursor image so that it matched the location of a target image. This
task forms a simplified framework by which task selection, as
expressed in limb movements and distinct from (or indeed, together
with~\cite{trumbo_motor_1968}) saccadic eye movement, can be
studied. However, dissociates the end-points of the saccade and
limb movements in a way which is not typical of natural behaviour; in
the example, the target of the animal's saccade to the blue berry
matches the target of the reach to grasp the berry. Studies which
investigate the positional reference frames employed by the brain for
saccadic and reach movements may not be expected to observe the same
results for the arm movements of the step-tracking tasks as for
natural reach-to-grasp movements.

Experiments to investigate the nature of arm movements are typically
laboratory based. Like the step-tracking experiments described above,
they usually consist of a screen to display the stimuli and an input
system which the subject uses to interact with the display. The input
may be via a set of
buttons~\cite{tipper_selective_1992,waszak_intention-based_2005}, a
joystick~\cite{pratt_action-centered_1994} or a computer
keyboard~\cite{elsner_effect_2001}. Movement trajectories are
collected using specialised
equipment~\cite{song_target_2008,jax_hand_2007} or may only be
inferred from information about the trajectory end
point~\cite{pratt_action-centered_1994}.

Here, addressing the issues of complexity, lab-bound constraint and
target/effector end-point dissociation, we present the \emph{line
  task}; a novel, inexpensive and easy to use experimental apparatus
for the study of one dimensional target selection and reach. The task
is designed to minimise the cognitive complexity of the target
selection and reach-tasks which it is intended to probe. Thus, it uses
vertical lines which move in one dimension only (horizontally),
meaning that subjects could plan their movements internally in a one
dimensional space, even though the end point moves in two dimensions
and the arm movement is carried out in multiple rotational dimensions.
Colour is used to allow subjects to distinguish between target and
distractor stimuli. No other object features need to be recognised by
the subject's visual processing systems. The apparatus provides the
ability to record reaction times and detailed trajectory
information. Because it is implemented on standard tablet computer
hardware, the line task is not constrained to the laboratory and can
be used by individuals in their own homes to generate movement data
which can then be collected via the internet for analysis by the
researcher.

The device was tested and evaluated on an action selection task (with
synchronous and asynchronous distractor) on a large sample size of 55
test subjects, to verify that the device met the design
requirements. The task was chosen to demonstrate the particular
utility of the device design; tracking a target line in the presence
of a distractor line. This task revealed for the first time a
particular deficit in action selection with an asynchronous
distractor: increased latency compared to a synchronous
distractor. This new result is distinct from previous action selection
investigations because the results do not have potential confounding
factors associated with higher level cognitive loading and explicit 3D
visual/movement transformations, illustrating the value of the line
task.

\section*{Materials and Methods}
\subsection*{Line task}

\begin{figure}[htb!]
\centering
\includegraphics[width=0.9\textwidth]{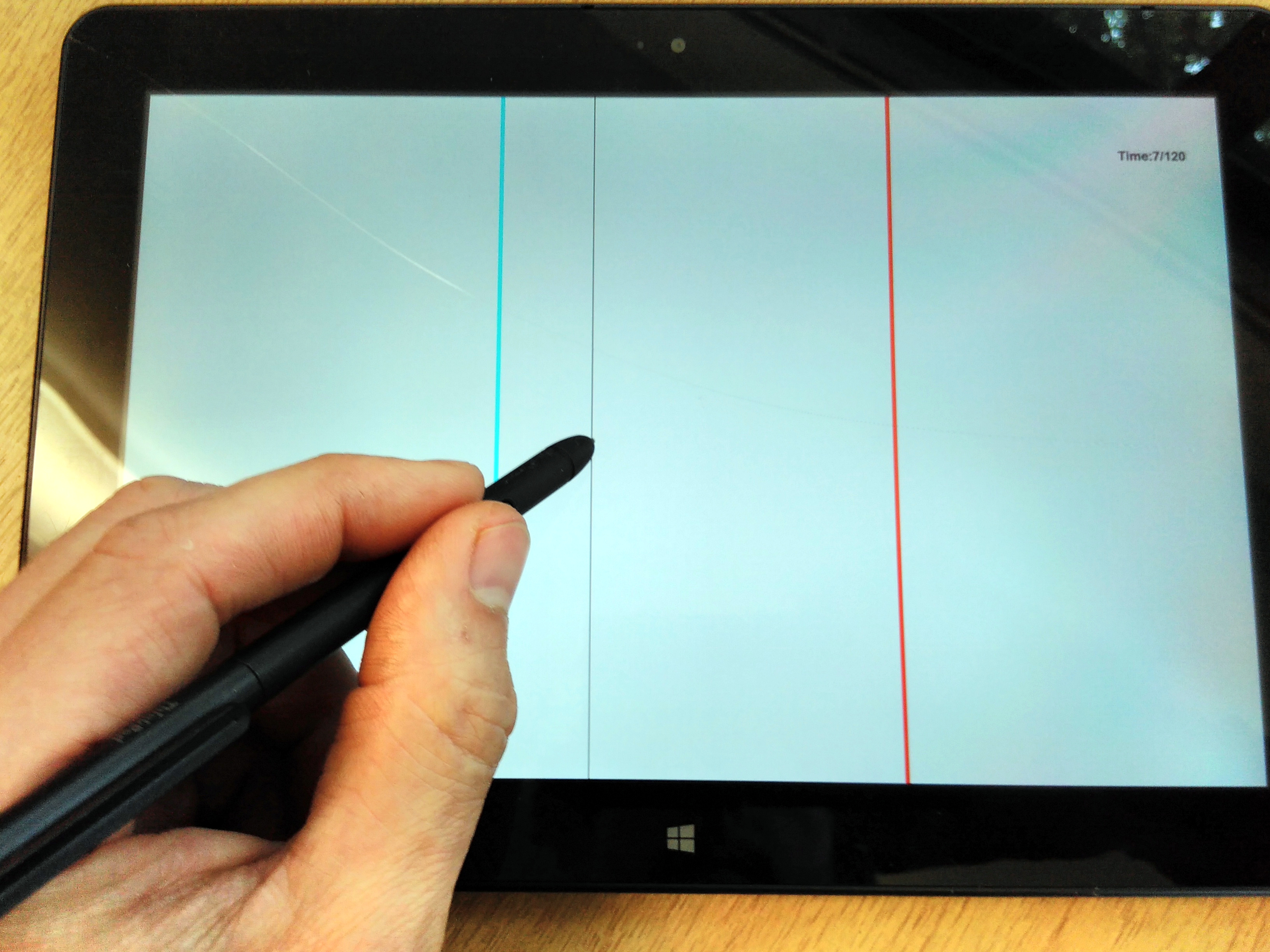}
\caption[The line task.]
{The line task in operation. The cyan
  target line is visible, along with a red distractor and the
  narrower, black stylus line.}
\label{linetask}
\end{figure}

The line task was presented on a tablet PC (see Fig~\ref{linetask}
and \cite{notremor_video_2015}). It requires participants to move a
stylus to the location of a vertical, cyan line (the target) whose
position changes randomly, possibly in the presence of a red line (the
distractor), whose position also changes. There were three
experimental conditions for the line task. One in which there was `No
Distractor' (ND), one in which the distractor changed position
asynchronously with respect to the target line (`Asynchronous
Distractor' or AD), and one in which the distractor line changed
location whenever the target's position changed (`Synchronous
Distractor' or SD). In each condition, the target line would spend a
randomized period of time $T_t$ at each location before simulaneously
disappearing from the old location and appearing at the new
location. $T_t$ is drawn from a normal distribution with mean 1.6~s
and variance 0.636:
%
% Rewriting:
% T_t = A + B N(\mu, \sigma^2) = A + N (B\mu, B^2\sigma^2) = N(A+B\mu, B^2\sigma^2)
%
% so for A=0.8, B=0.005, \mu=160, \sigma^2=25440: T_t = N (\mu'=1.6, \sigma'^2=0.636)
\begin{equation}\label{eq:tau_target}
  T_t \sim N(\mu=1.6,\sigma^2=0.636)~\text{seconds}
\end{equation}
%
% The number of timesteps which the program waits in addition to the
% 0.8 seconds is produced by the following method:
%
% Loop. In each loop, draw a number from a uniform distribution (0 to 1) and
% compare with the number p = Timestep(s) / 0.8 = 0.005/0.8 = 0.00625
% (p is called lambda in the actual Windows Line task code). If the
% random number is less than p, then change line position.
%
% This is an example of a Bernoulli trial with p = 0.00625 and
% q = 1-p = 0.99375.
%
% The probability mass function here is the geometric function:
%
% Pr (X=k) = (1-p)^(k-1) . p
%
% This gives the probability of having k-1 failures before success on
% the kth try (or timestep as in our example). The mean of the
% resulting distribution is 1/p = 160 and the variance of the function is
% (1-p)/p^2 = 25440 meaning that the number of timesteps which will be waited
% for will be drawn from a normal distribution with mean 160 and
% variance 25440.
%
In the synchronous distractor condition, the time between distractor
location changes is identical to $T_t$. The new target location, $x$,
was computed by first generating a random location offset,
${\delta}x$, then adding this to the current location, subject to the
bounds of the edge of the screen:
% See line 212 of line.cpp (of the Windows line task) for the code
% which produces the following. Note that the final tests in the cpp
% code (lines 215 & 216) are not necessary.
\begin{equation}\label{eq:rand_loc_offset_sync}
  {\delta}x \sim \frac{w}{4}~U(-1,1)
\end{equation}
\begin{equation}\label{eq:new_loc_sync}
  x=
  \begin{cases}
    x' + {\delta}x & \text{for }\frac{w}{10}<x'+{\delta}x<\frac{9w}{10}\\
    x' - {\delta}x & \text{otherwise}\\
  \end{cases}
\end{equation}
where $x'$ is the previous location of the target, $w$ is the width of
the screen in pixels and $U(-1,1)$ is a uniformly distributed random
32 bit floating point number between -1 and 1.  In the asynchronous
distractor condition, the time between target location changes is the
same as in the ND and SD condition with $T_t \sim
N(\mu=1.6,\sigma^2=0.636)~\text{s}$. The time between distractor
location changes was given by:
%
% Here, p = Timestep(s) / 0.4 = 0.0125 so the normal distribution for
% the jumps has mean 80 and variance 6320.
%
% Rewriting this:
% T_d = A + B N(\mu, \sigma^2) = A + N (B\mu, B^2\sigma^2) = N(A+B\mu, B^2\sigma^2)
%
% so for A=0.4, B=0.005, \mu=80, \sigma^2=6320: T_d = N (\mu'=0.8, \sigma'^2=0.158)
%
\begin{equation}\label{eq:tau_distractor_async}
  T_d \sim N(\mu=0.8,\sigma^2=0.158)~\text{seconds}
\end{equation}
Note the shorter periods; the asynchronous distractor changes location
more quickly than the target.
The locations of the asynchronous distractor are calculated according to:
\begin{equation}\label{eq:rand_loc_offset_async}
  {\delta}x \sim \frac{w}{2}~U(-1,1)
\end{equation}
\begin{equation}\label{eq:new_loc_async}
  x=
  \begin{cases}
    x' + {\delta}x & \text{for }\frac{w}{10}<x'+{\delta}x<\frac{9w}{10}\\
    x' - {\delta}x & \text{otherwise}\\
  \end{cases}
\end{equation}
which differs from the computation of the target locations only in
that the distractor location is permitted to change by up to half the
screen width, which allows the distractor to appear either side of the
next target location, even if that is at the most distal location
possible.  In the synchronous distractor condition, the distractor
line changes location simultaneously with the target line, and by an
equal and opposite distance. If the distractor line would need to be
displayed beyond the edge of the screen, then it was placed $w/20$
pixels from the edge.

\subsection*{Experimental design}
The line task experiment had a one way, repeated measures design. The
dependent variables were the movement latency and the error
rate. Movement latency refers to the time taken to initiate a movement
after target stimulus onset. Significant movement away from the target
counted as an error. The definitions of error and latency are given in
\emph{Latency extraction}, below. The related samples independent
variable was the distractor condition, with three levels (synchronous,
asynchronous and no-distractor). A linear mixed effects model was
fitted with distractor condition as the fixed effect and individual
included as the random effects component. An alpha value of 0.05 was
selected for significance testing.

\subsection*{Apparatus and stimuli}
The line task was presented on a Lenovo ThinkPad 10 (20C1) tablet
computer with a 218 mm x 137 mm (1920 pixel x 1200 pixel) colour
display. The tablet was positioned in landscape orientation and thus
the height was 137 mm and the width was 218 mm (see
Fig~\ref{linetask}). A vertical cyan (RGB24 representation: \#fcfc00)
target line was displayed on the white (\#fcfcfc) background of the
display, along with a vertical black (\#000000) line indicating the
current horizontal position of the participant's stylus (the vertical
location of the stylus was not recorded or displayed). In some cases,
a third, vertical red (\#fc0000) distractor line was also
displayed. All lines extended from the top of the display to the
bottom; the full height of the display. Target and distractor lines
were 8 pixels wide; the black line indicating current position was 2
pixels wide and overlaid the target or distractor lines when
co-located with either of them. A stylus (ThinkPad Tablet Digitizer
Pen, manufacturer: Wacom) was used to track the target line. Use of
this particular stylus disables the capacitive touch screen on the
ThinkPad, preventing contact with the heel of the hand from
registering as an erroneous stylus position. The position of the
stylus was digitised approximately every 5~ms.
A software timer guaranteed that the stylus position was recorded at
least every 5~ms, though if the system was busy, a longer gap could
occur between recordings. For each datum, the time was recorded with a
precision of 1~ms.
The refresh rate of the tablet display screen was 60 s$^{-1}$. The line
task was programmed in C++ using the Qt framework version 5.3.

\subsection*{Participants}
Overall, 61 psychology undergraduate students participated in the
study. 6 participants were removed prior to analysis due to errors in
the implementation of the tasks (the target's mean jump time had been
set to values other than 1.2 s for one or more conditions). This left
55 participants (9 males and 46 females). 47 of these participants
were right handed; 8 were left handed. All participants
had normal to corrected vision. The chronological age of the 55
participants ranged from 18 to 28 years old with a mean age of 19.2
years (SD=2.1).

The majority of participants were an opportunity sample. These
participants were recruited via the University of Sheffield's online
research participation system, in order to gain course credit.  The
experiment was conducted by five different experimenters, who were all
involved in the recruitment and data collection. The study received
ethical approval from the Psychology department's ethics
sub-committee. Written informed consent was also collected from all
participants.

\subsection*{Procedure}
Each participant completed the line tracking tasks individually within
the same dimly lit cubicle room.
On arrival, the participants were asked to sign a consent form which
outlined the experimental aim. They were then given verbal
instructions to follow the blue target line with the stylus pen in all
trials, while ignoring other stimuli. Participants were explicitly
told that they must complete the tasks ``as quickly and as accurately
as possible'', implicitly allowing each participant a decision
between movement speed and accuracy.
The tablet was placed
flat on the table, directly in front of the participant. The tablet's
display brightness setting was set to automatically adjust based on
the surrounding light level. To start the tasks, the participants were
verbally instructed to press the play button with the stylus pen when
they were ready. Participants then completed a 30 second practice
% Katie wrote a mean jump time of 1.5s which is what the UI said BUT
% this means 1.5 + 0.005 N(300,89700). The UI was mislabelled.
%
% Here, p = Timestep(s) / 1.5 = 0.00333333 so the normal distribution for
% the jumps has mean 300 and variance 89700. and the mean jump time is
% 1.5 + 1.5 = 3 seconds.
%                                -----------v
task in which the mean jump time was set to 3 seconds (thus slower
than the main tasks during which data was recorded). After the
practice task, participants had a chance to ask any further questions
about the tasks, after which they proceeded to complete the three task
conditions (ND, SD and AD), each of which lasted for 120
seconds. Before the participant started the tasks, the experimenter
would leave the room and dim the lights.
The sequence in which the task conditions were presented was
determined randomly for each participant. When participants had
completed the first task, they informed the experimenter. The end of a
task was indicated by a return to the line task application's start-up
screen. The experimenter would then enter the room and adjust the
settings on the tablet for the next task. The participant would then
proceed to complete the next 120 second task condition. This process
was repeated until the participant had completed all three task
conditions.

\subsection*{Latency extraction}
The data collected, and all of the scripts used to analyse them are
available at \url{https://github.com/ABRG-models/linetask2014}.

The raw data generated by the line task application consisted of
position data for the target line, the distractor line (if any) and
the stylus position, sampled at approximately 5~ms intervals. For each
subject, one raw data output file was produced for each of the
experimental conditions.  These raw data files were analysed by a
script (\filename{lt\_analyse.m}) implemented in Octave (version
3.8.2), which algorithmically recognises the beginning and end of a
stylus movement, returning latency to first movement for each trial,
along with movement error information. The volume of data necessitated
an automated approach; there were a total of 17617 individual target
and distractor events to analyse. In order to verify that the latency
extraction script was producing reliable results, an alternative
method based on resampling and movement filtering was also applied to
the ND and SD cases, although this method was not extended to handle
the many subtleties of the more complex AD condition.

% This describes script1 - the Octave script.
\paragraph{Primary latency extraction method.} The first latency extraction
method was based on the raw stylus position information and its first
derivative.

\begin{figure}[htb!]
\centering
\includegraphics[width=1\textwidth]{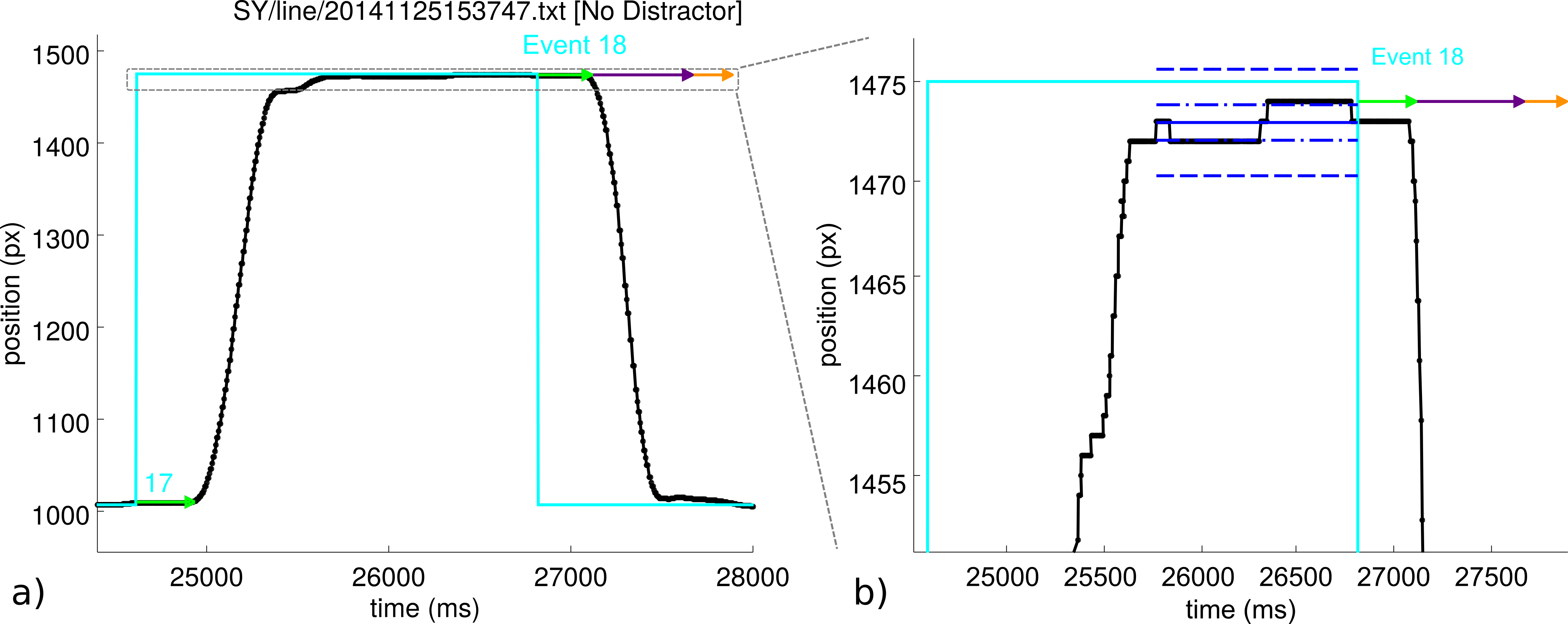}
\caption[Obtaining the latency, method 1] {\textbf{a)} An example
  event --- number 18 from this data set. The horizontal position of
  the vertical target and stylus lines is plotted against time. The
  black line shows the stylus position. The cyan line shows the
  position of the target. At event 18, the position of the target
  shifts from 1475 px to 1007 px. Event 17 is also visible. The grey
  dashed box indicates the region of the graph which is expanded in
  plot b). \textbf{b)} The expanded region from plot a) which shows
  the result of the movement onset analysis in more detail. Blue lines
  indicate the `stable stylus position' period which precedes the
  target event. Blue dot-dash lines indicate 1 standard deviation in
  position over this period; blue dashed lines indicate 3 standard
  deviations. The latency to first movement is shown as a green arrow,
  with a purple arrow showing the algorithm's estimate of the length
  of the first, uninterrupted smooth motion and the orange arrow
  indicating the detected time for the motion to complete. Here, the
  subject moved in the correct direction.}
\label{example_event}
\end{figure}

% Put what the script does in present tense because it still exists
% and still operates like this.
The script begins by identifying the times of target and distractor
events. A target/distractor event is the event that the
target/distractor changes position on the screen. To determine the
latency to movement for a given target event it is necessary to find
the time at which the stylus first begins to move. The script computes
the average position of the stylus prior to the event; the `stable
stylus position'. The mean and standard deviation of the stylus
position data from the time of the end of the previous movement to the
time of the current target event is calculated.  Motion is detected
if, post-event, the stylus position departs further than 3 pre-event
standard deviations from the pre-event mean position and has a speed
% px/ms times 1000 gives px/s
% px/ms times 11.4 gives cm/s
greater than 0.05 pixels per millisecond (50~px/s; 5.7~mm/s,
cf.~\cite{prablanc_optimal_1979}).  The displacement at this time step
gives the initial stylus direction. To determine as closely as
possible the start of the movement, the algorithm steps back along the
trajectory until the stylus position is within 1 pre-event standard
deviation from the pre-event mean position (dash-dot blue lines in
Fig~\ref{example_event}). This gives the movement latency, shown in
the figures as a green arrow.

% Use event 18 of AllData/Jon/SY/line/20141125153747.txt (ND)

After a target event, if the motion was initially in the direction of
the target line's new position, then this was recorded as a correct
movement. If the motion was in the opposite direction, this was
recorded as a movement error. Feedback on whether or not the movement
was in error was \emph{not} given to the subject during the task. Note
that a very brief movement in the wrong direction was counted as an
error; in most cases subjects correct their movements quickly (in
around 100~ms). A motion towards the distractor line following a
distractor event was also recorded as a movement error. In the
asynchronous condition, a distinction was made in the analysis between
target and distractor movement errors.

It was not possible to measure a latency for every event; some were
omitted from this analysis. Most of these events were also omitted
from error analysis.
% The exceptions are the distractor events for
% which the stylus was NOT distracted. There was
% therefore no error, but no latency could be
% measured either!
A list of the reasons for omitting events is given in S1 Appendix.

The analysis of events was significantly more complex in the
asynchronous distractor condition. In both the no-distractor condition
and the synchronous distractor condition, each event was well
separated from preceding and subsequent events. In most cases, as long
as a stable stylus position was achieved before the event, then the
latency and the movement direction error could be easily determined,
as in Fig~\ref{example_event}. The difficulty in the asynchronous
distractor condition was that distractor events could occur at any
time, and had been programmed to occur in larger numbers, so that
instead of analysing 50 to 60 target events (or target+distractor
combined events for the synchronous condition) it was necessary to
analyse roughly 50 to 60 target events and also over 100 distractor
events for each trial.

\begin{figure}[htb!]
\centering
\includegraphics[width=0.5\textwidth]{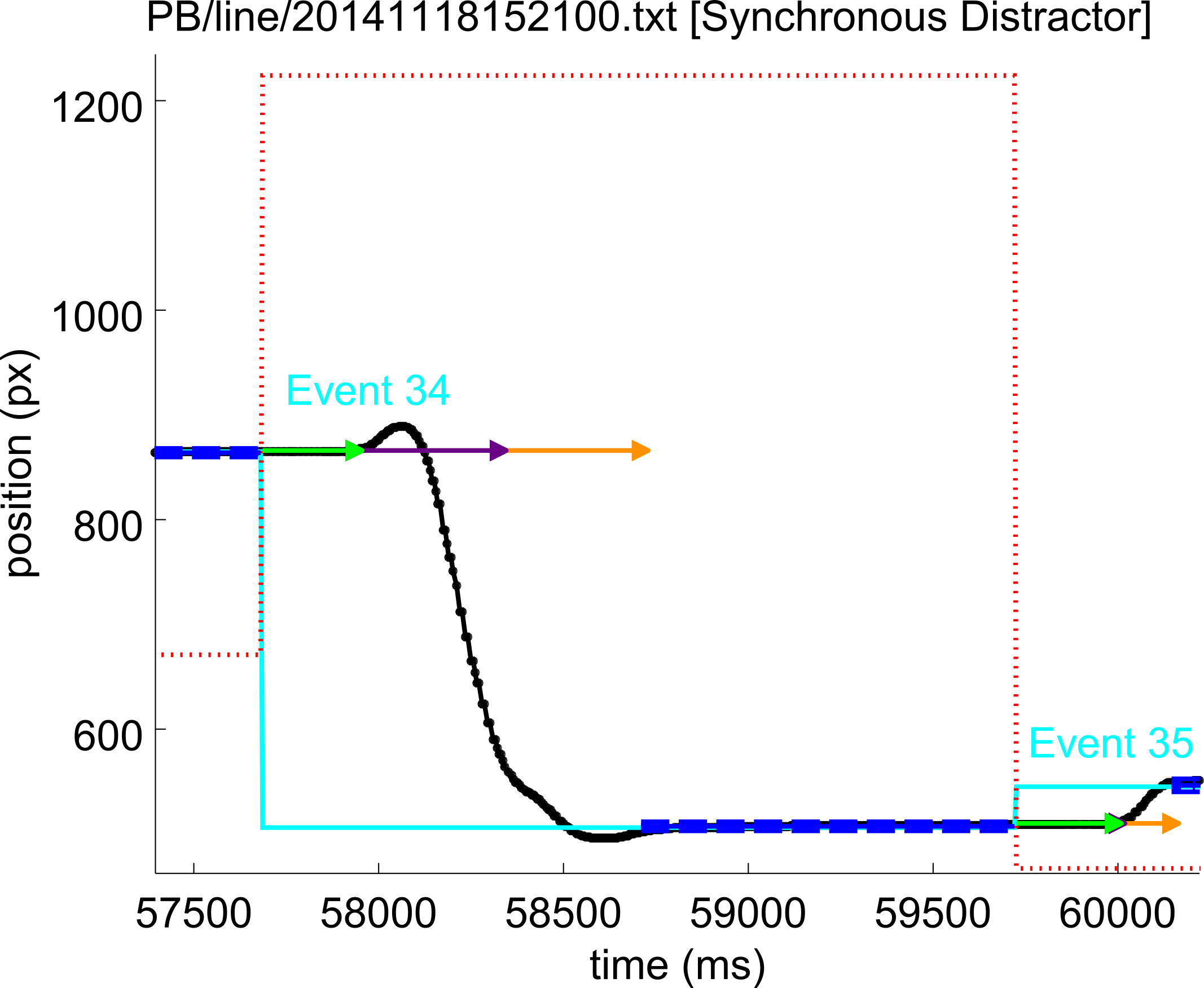}
\caption[Example error event] {An example movement error event. The
  line colours have the meanings given in
  Fig~\ref{example_event}. The red dotted line is the horizontal
  position of the vertical red distractor line. Following event 34,
  this subject moved the stylus towards the distractor (red dotted
  line). Within 100~ms of the movement, the subject corrected the
  direction and moved to the target position.}
\label{example_errorevent}
\end{figure}

% Error rates
\paragraph{Movement errors.} In the statistical analysis for the no
distractor condition, movement errors were always associated with
target events. In the synchronous condition, because a distractor
event always occurred simultaneously with the target event, movement
errors could have been associated with either event; the target event
was chosen to match the no-distractor condition. For the
asynchronous distractor condition, movement errors which correlate
with target events were distinguished from those which followed
distractor events. For a target event, a trajectory which was initially
directed \emph{away} from the target was recorded as a movement error. For a
distractor event, a movement \emph{towards} the distractor was considered to
be a movement error. Fig~\ref{example_errorevent} shows an example movement
error event.

% Rachel/PB/line/...2100.txt

\paragraph{Alternative latency extraction method.}

To verify the accuracy of the latency extraction described above, an
alternative approach was developed, and the results compared for the
no-distractor and synchronous distractor conditions.

% Mauro's method contribution
This method relies on the computation of the sign of the stylus
velocity. Because the touch position is quantized, the stylus velocity
is exactly zero as long as the readout position is the same pixel. If
the stylus drifts slowly then a series of short pulses are observed in
the velocity. Voluntary movements are sustained for some time and
hence, the velocity maintains the same sign for the same interval.

\begin{figure}[htb!]
\centering
\includegraphics[width=0.7\textwidth]{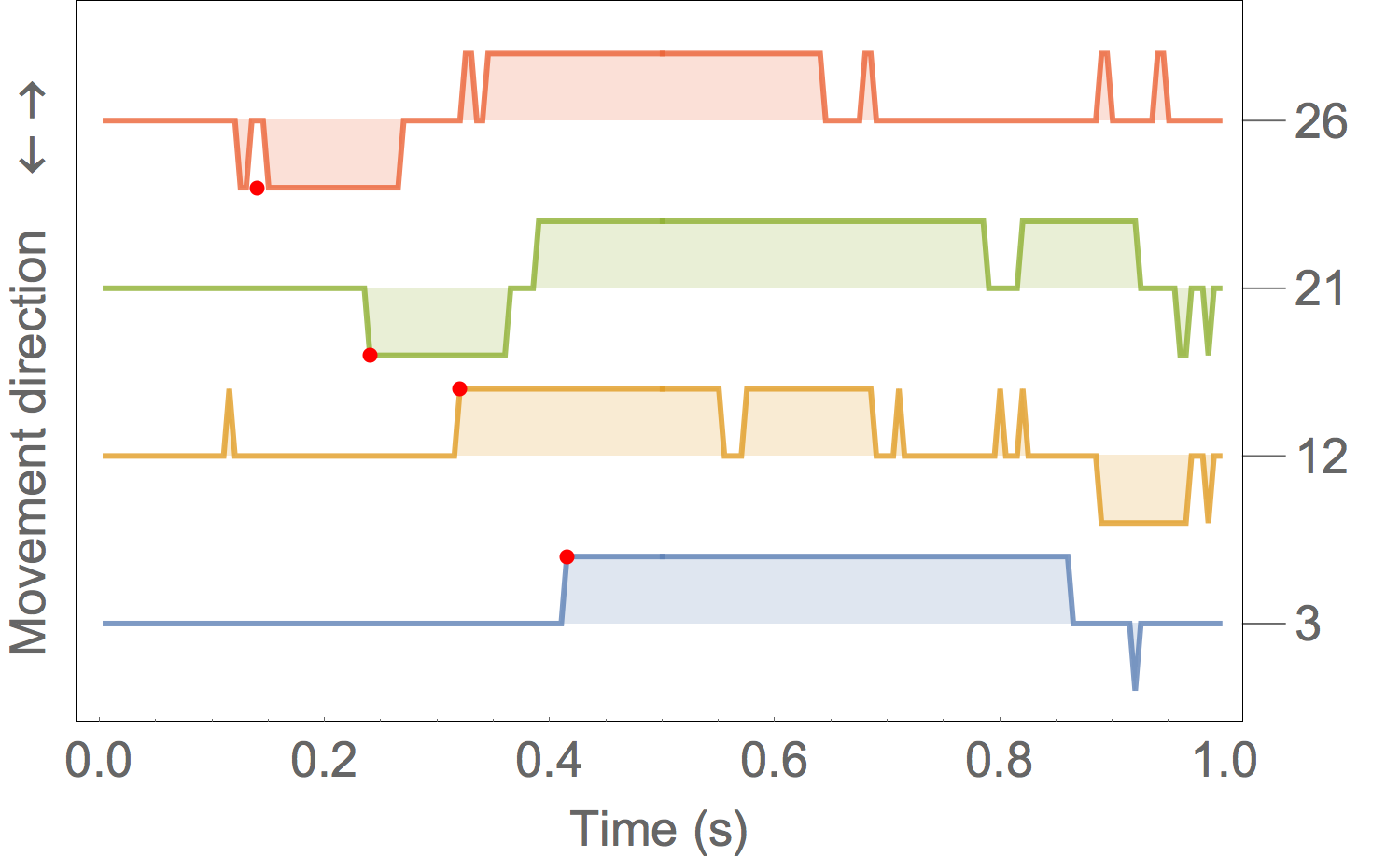}
\caption[Alternative method] {The sign of the stylus velocity plotted
  for 4 example events, numbered 3, 12, 21 and 26 on the right hand
  axis. Each event is aligned with the target position change at 0.}
\label{alt_method_move_dirn}
\end{figure}

Fig~\ref{alt_method_move_dirn} gives an example. In the event 3, the
stylus remained stationary over the same pixel until time
t=0.4~s. Then the sign of the velocity remains positive (towards the
target) until t=0.85~s. This is an intentional target reach
movement. Note that at about t=0.9~s there is a negative velocity spike
which represents the stylus stepping back one pixel. Event 12 is more
complex; in this case at about t=0.1~s there is a spike because the
stylus steps to the next pixel. If spikes are ignored, the real
movement is made of three steps two initial steps towards the target
and one backwards (because the target was overshot). Example 21
represents an error: at t=0.25~s the stylus moves away from target,
but then, at t=0.38~s the direction of movement is correct towards the
target. Example 26 is another error example: in this case the movement
starts slowly in the wrong direction and looks like a spike preceding
the sustained phase.

In order to detect the time when an intentional movement begins,
spikes needs to be ignored. This has been done by filtering the
velocity sign signal with a median filter (the radius of the median
filter determines the duration of the neglected spikes and was 40~ms
in our implementation, which also means that only movements lasting
longer are considered intentional). The red dots in
Fig~\ref{alt_method_move_dirn} represent the detected movement
onset. The sign of the velocity at that point indicates whether the
movement is correct or an error.

The method may produce outliers of two kinds: a) if there is no
movement, no delay is computed (a condition that had instead to be
explicitly omitted above), b) if the subject is distracted and reacts
too late, a long reaction time is produced which is filtered as an
outlier.

This method has been used with only two conditions: a) target
jump size greater than 20 pixels (as above) and b) drift of stylus
smaller or equal to 1 pixel in the first 0.15~s. These two conditions
are intentionally different from those listed above so that we can
compare the results of a different method with slightly different
omission criteria.

This second method has confirmed the estimation carried out with the
primary method: the mean reaction time with this method is
systematically slightly shorter than the primary method due to
differing criteria for event omission. Table.~\ref{table:twosample}
includes the results for the alternative method alongside those for
the primary method.
% End of Mauro's method contribution.

% Do I want to incorporate an analysis of the initial movement
% distance? Not really - it's already complicated enough.
\subsection*{Statistical analysis}

% The detail here to go in a readme:
%
%The task parameters and the latency and error data generated by the
%Octave script were stored in a single file named \filename{fnames.mat}
%(MATLAB v.~7 binary format). The ipython notebook
%\filename{Accessing\ Data.ipynb} describes the content of the data
%structures in \filename{fnames.mat}. An ipython notebook called
%\filename{Anova.ipynb} was used to apply the various statistical
%analyses (using python version 2.7.6) which are reported here. Some
%analyses were carried out in R (version 3.0.2)~\cite{r_core_team_r:_2013}, in
%which cases, \filename{Anova.ipynb} made calls to R, displaying the
%results within the notebook.
The statistical analyses reported here were carried out in
R~\cite{r_core_team_r:_2013} and python.

\begin{figure}[htb!]
\centering
\includegraphics[width=0.5\textwidth]{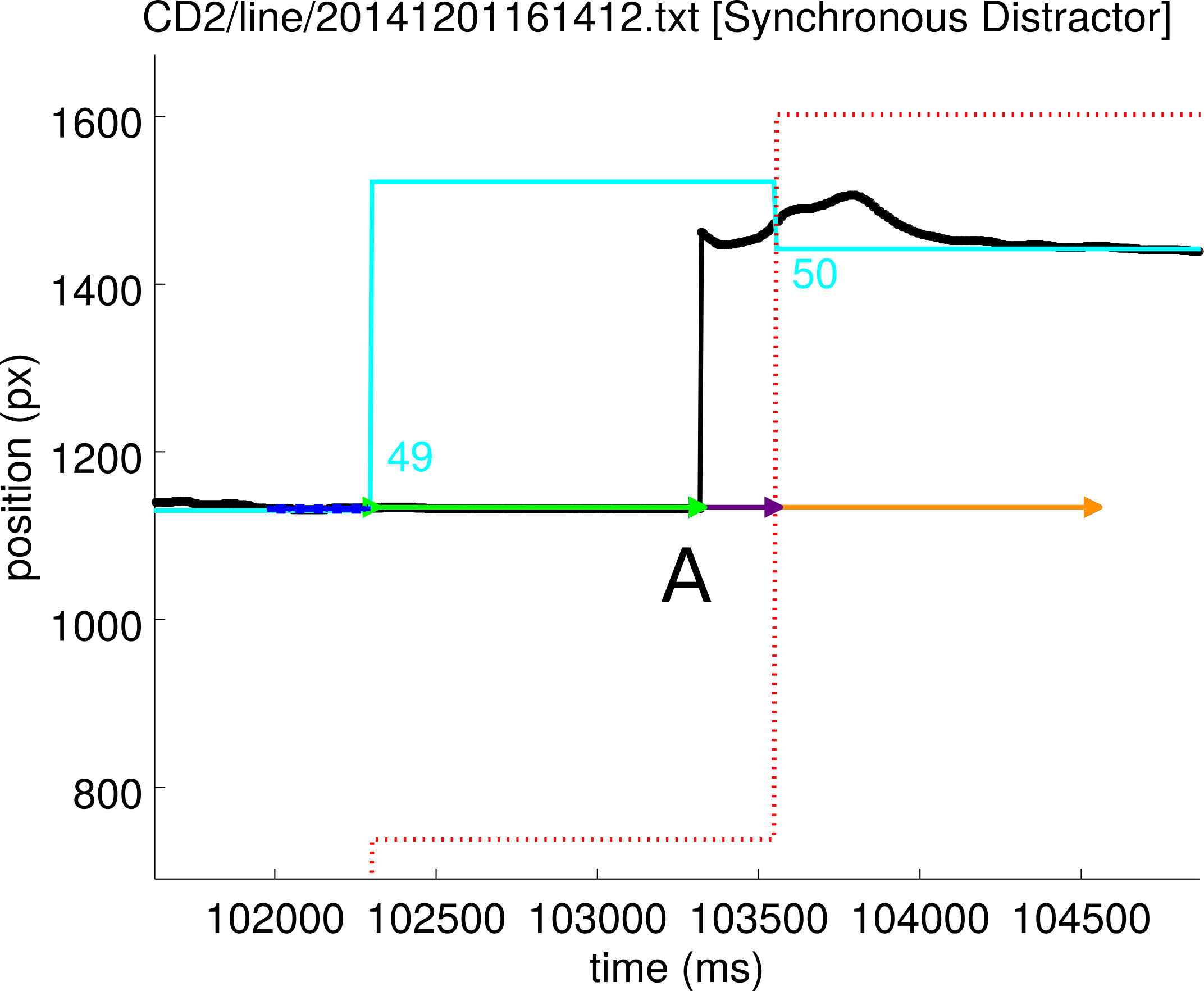}
\caption[Stylus lifted] {An example event during which the stylus
  (black line) lifted and it seems that movement did not occur until
  1013~ms after the target event (label `A') when the stylus made
  contact with the screen and data could again be recorded. This
  latency was recorded by the latency extraction script, but was then
  excluded by the median absolute deviation method employed to exclude
  outliers. The cyan line is the target position, the red dotted line
  is the distractor position. For the meaning of other coloured lines,
  refer to Figs~\ref{example_event}~and~\ref{example_errorevent}}
\label{stylus_lifted}
\end{figure}

Prior to applying analyses, outlier data were excluded from the
latency measurements for each individual. For details, code and data,
see S2 Appendix and \url{https://github.com/ABRG-models/linetask2014}.
Causes of outlier data included an individual becoming distracted from
the task and missing one or more targets or the stylus lifting from
the screen and the position failing to update, which occurred for the
example shown in Fig~\ref{stylus_lifted}. Outliers were excluded
using a ``median absolute deviation from the median'' method
\cite{boris_iglewicz_how_1993} with the modified Z-score threshold set
to 3.5. After excluding outliers, the mean latency in each condition
was computed for each individual.

% Example here is event 49 of Rachel/CD2/line/20141201161412.txt

% A nice example of a simply very long reaction is Event 180 of
% Olivia/BG/line/20141120153210.txt. Figure not shown for now.

\section*{Results}

Other than where indictated, the results presented here have been
derived from analysis of latencies and error rates produced by the
primary latency extraction method.

\subsection*{Latency extraction}

The mean number of target events in the tasks was 64.7 with standard
deviation (SD) of 6.9. Of these, the Octave latency extraction script
was able to measure latency to first movement in 53.1 (SD=7.8)
cases. The script was most successful in collecting latencies from the
no-distractor task condition, returning a latency for 89\% of trials
across individuals (range 75\% to 100\%). The presence of a distractor
line increased the variability of the stylus trajectory, reducing the
number of target events for which a latency could be determined. In
the synchronous distractor condition, latencies were returned from
79.1\% of trials (range 58\% to 96\%); in the asynchronous condition
from 78\% of trials (range 57\% to 100\%).

\subsection*{Analysis of latencies}

\begin{figure}[htb!]
\centering
\includegraphics[width=0.7\textwidth]{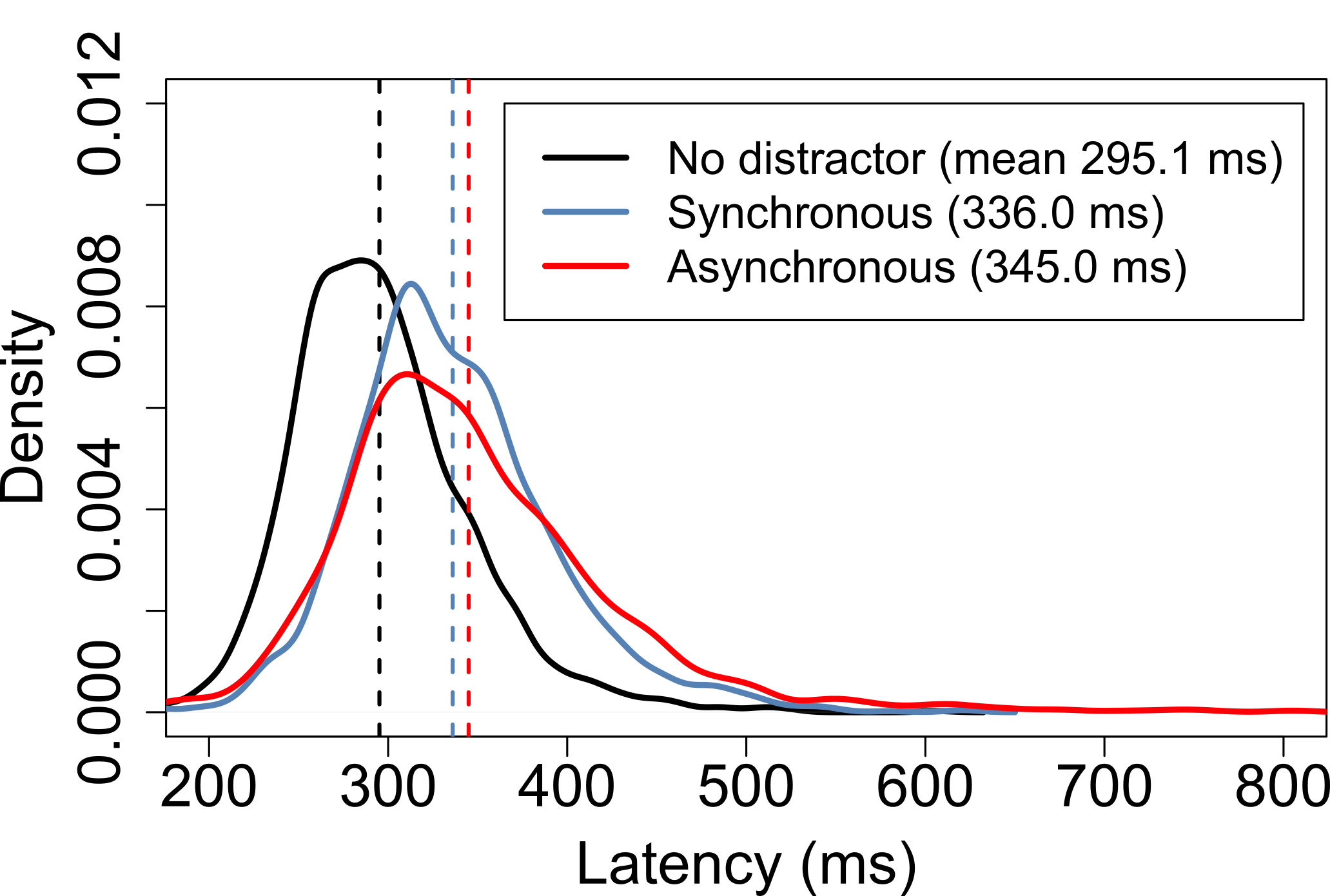}
\caption[Latency measurement PDF] {Probability density function
  approximations for 2013 latency measurements for each of three
  experimental conditions. The graph indicates the wide variances in
  movement onset latencies which are typical in any animal behavioural
  experiment. This graph shows latencies for correct movements
  only. Black indicates data for the no distractor condition, blue for
  the synchronous distractor condition and red for the asynchronous
  distractor. Dashed lines indicate the means. Although the
  synchronous and asynchronous curves appear very similar, bootstrap
  resampling of the two samples indicates that there is a very low
  probability that they are generated by the same population
  distribution.}
\label{data_density}
\end{figure}

Fig~\ref{data_density} shows probability density functions for a
subset of the measurements taken from all 55 subjects in the three
conditions. Here, the maximum possible equal number of
correct-movement latency measurements from each condition were
randomly sampled for each subject, giving datasets of length 2013 for
each condition.

\begin{figure}[htb!]
\centering
\includegraphics[width=0.7\textwidth]{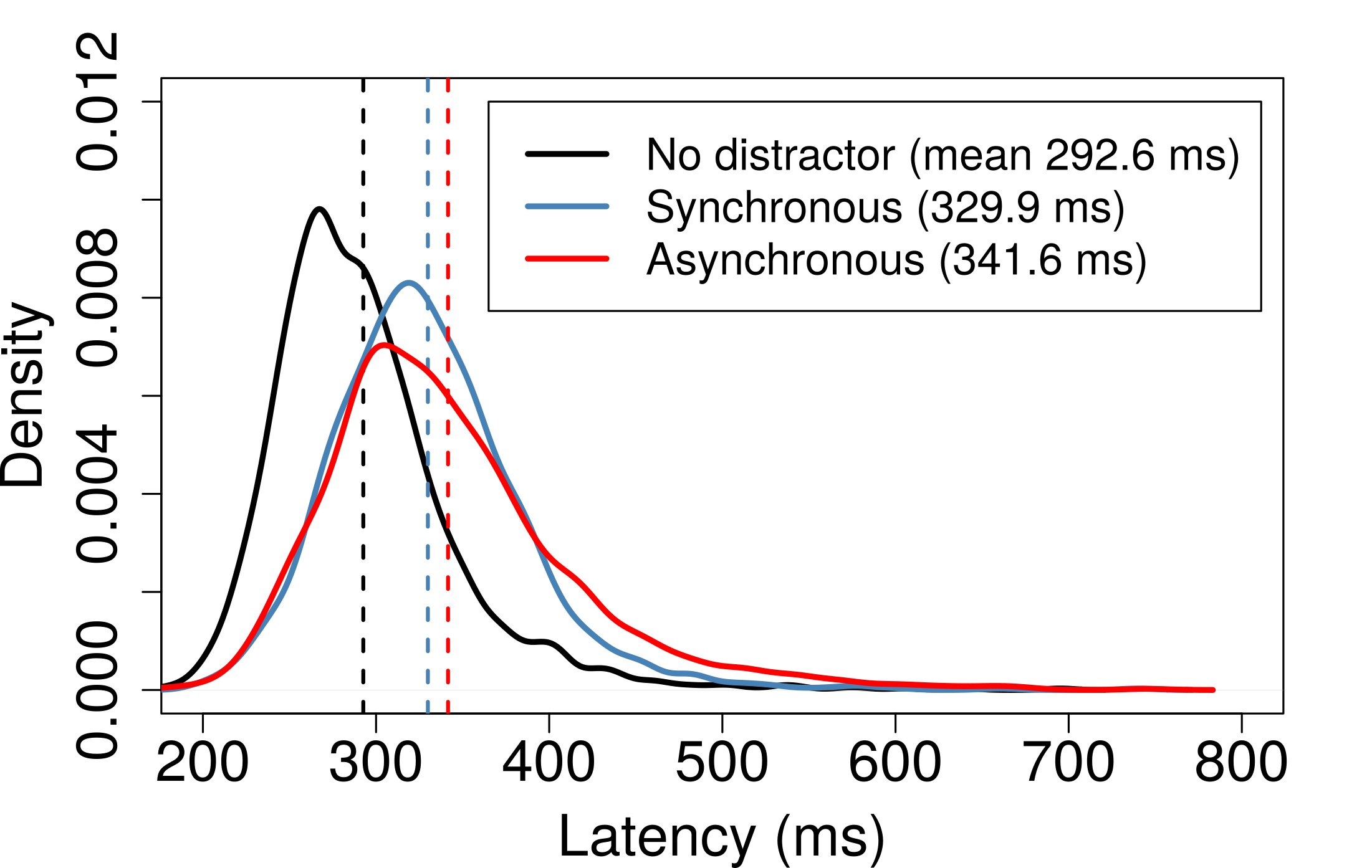}
\caption[Latency measurement PDF] {Probability density function
  approximations for latency measurements determined using the
  alternative method. The presentation matches Fig~\ref{data_density}:
  Black indicates data for the no distractor condition, blue for the
  synchronous distractor condition and red for the asynchronous
  distractor. Dashed lines indicate the means.}
\label{data_density_alt}
\end{figure}

Fig~\ref{data_density_alt} shows the result of the same analysis of
the latencies determined by the alternative latency extraction
method. Inspection of Fig~\ref{data_density_alt} shows that the
latency distributions for the alternative extraction method preserve
the form of those shown in Fig~\ref{data_density} for the primary
method. The sample means and standard deviations are summarised in
Table~\ref{table:latmeans}. To estimate the standard error of the
sample mean, we created bootstrap samples of 2013 observations drawn
with replacement and then computed the mean for each of
4$\times$10$^6$ replications.

\begin{table}[ht]
\caption{Summary of latency means with standard deviations, for the
  primary and alternative latency extraction methods. A bootstrap
  estimate of the standard error of the mean is also given.}
\centering
\begin{tabular}{c c c c}
\hline
\textbf{Condition} & \textbf{Mean (ms)} & \textbf{Std. Dev.} & \textbf{Std. Err.} \\ [0.5ex]
\hline
~ & \multicolumn{3}{c}{Primary latency extraction method} \\
\hline
ND & 295.1 & 51 & 1.1 \\
SD & 336.0 & 57 & 1.3 \\
AD & 345.0 & 79 & 1.8 \\ [1ex]
\hline
~ & \multicolumn{3}{c}{Alternative latency extraction method} \\
\hline
ND & 292.6 & 53 & 1.3 \\
SD & 329.9 & 56 & 1.5 \\
AD & 341.6 & 71 & 2.0 \\ [1ex]
\hline
\end{tabular}
\label{table:latmeans}
\end{table}

%Normality of data:
% Shapiro-Wilks:
% No Distractor, 25 sub-samples: W 0.914870500565 p-value 0.0391994863749 (Reject Normal NULL hypothesis if p<alpha=0.05)
% Sync Distractor, 25 sub-samples: W 0.845669388771 p-value 0.00145912088919
% Async Distractor, 25 sub-samples: W 0.962160229683 p-value 0.459158211946

\begin{figure}[htb!]
\centering
\includegraphics[width=0.7\textwidth]{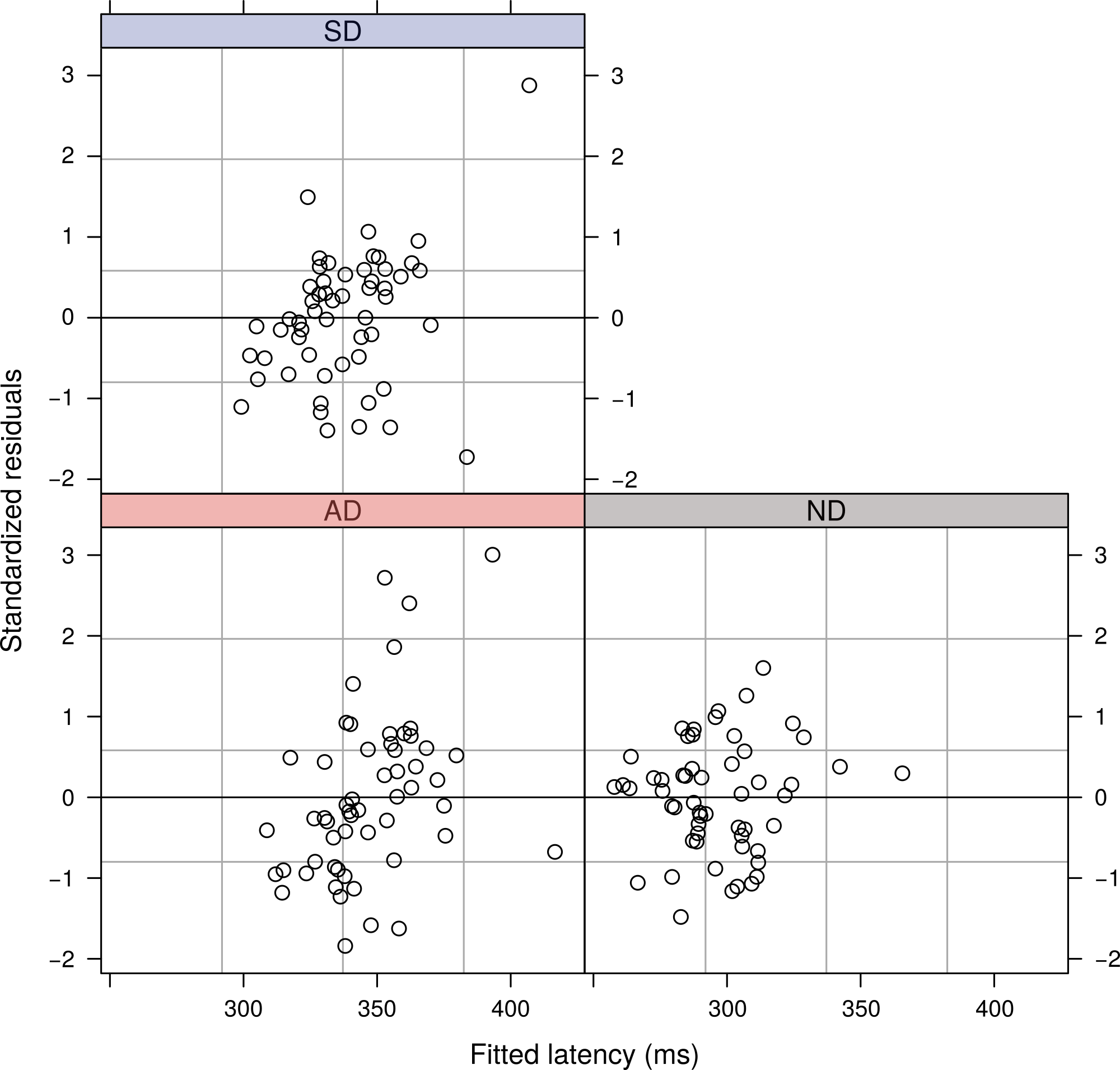}
\caption[Linear mixed effects model residuals] {Standardized residuals
  indicating the proximity of each individual subject's mean latency
  in the three conditions from the fit of the linear mixed effects
  model.}
\label{residuals}
\end{figure}

We constructed a linear mixed effects model on the latencies to
determine whether there was a statistically significant difference in
latency to first movement predicted by the fixed factor of the three
conditions `No Distractor', `Synchronous Distractor' and `Asynchronous
Distractor'. The individual subject was used as the random factor to
give an model of the form (using R-like notation):
\begin{equation}\label{eq:linear_mixed_effects_model}
\text{latency} \sim \text{condition} + (1|\text{subject}) + \epsilon
\end{equation}
%
%% > source ('Nonranked.r', print.eval=TRUE)
%% Linear mixed-effects model fit by REML
%%  Data: latdat
%%        AIC      BIC    logLik
%%   1513.105 1528.542 -751.5523
%%
%% Random effects:
%%  Formula: ~1 | subj_id
%%         (Intercept) Residual
%% StdDev:    22.41507 18.10252
%%
%% Fixed effects: latency ~ condition_str
%%                    Value Std.Error  DF   t-value p-value
%% (Intercept)     347.1905  3.885022 108  89.36643  0.0000
%% condition_strND -50.9244  3.452015 108 -14.75209  0.0000
%% condition_strSD  -9.5913  3.452015 108  -2.77847  0.0064
%%  Correlation:
%%                 (Intr) cnd_ND
%% condition_strND -0.444
%% condition_strSD -0.444  0.500
%%
%% Standardized Within-Group Residuals:
%%         Min          Q1         Med          Q3         Max
%% -1.84077745 -0.53695422 -0.01665622  0.51861454  3.00485476
%%
%% Number of Observations: 165
%% Number of Groups: 55
%%               numDF denDF  F-value p-value
%% (Intercept)       1   108 9615.924  <.0001
%% condition_str     2   108  122.904  <.0001
%%Data: latdat
%%Models:
%%nonranked2.null: latency ~ 1 + (1 | subj_id)
%%nonranked2.mdl: latency ~ condition_str + (1 | subj_id)
%%                Df    AIC    BIC  logLik deviance  Chisq Chi Df Pr(>Chisq)
%%nonranked2.null  3 1652.2 1661.5 -823.10   1646.2
%%nonranked2.mdl   5 1525.7 1541.2 -757.83   1515.7 130.53      2  < 2.2e-16 ***
%%---
%%Signif. codes:  0 ‘***’ 0.001 ‘**’ 0.01 ‘*’ 0.05 ‘.’ 0.1 ‘ ’ 1
%
The model showed that the different task conditions elicited
statistically significant changes in latency to first movement
($\chi^2$(2)=130.53, p\textless2.2$\times$10$^{-16}$). Plots of the
residuals from the model (Fig~\ref{residuals}) do not suggest any
evidence for heteroscedasticity or non-normality.

To determine differences and confidence intervals between pairs of
conditions we applied a bootstrap analysis to the data. We computed an
estimate of the standard error of the difference between the means in
the three conditions. These results are summarised in
Table~\ref{table:twosample}. The Z-scores indicate that the effect
size is large for either of the distractor conditions compared with
the no distractor condition, but small for the SD/AD pair.

\begin{table}[ht]
\caption{Summary of bootstrapped two-sample analyses of latency
  values. The difference between the sample means is shown (e.g.~for
  ND/SD, the SD condition has a latency which is 40.9 ms longer than
  for ND) along with the computed estimate of the standard error of
  the difference and the corresponding Z-score. The last column shows
  the result of the Studentized t-test to determine the probability
  that the two samples are drawn from the same population. For
  comparison, the corresponding results for the alternative latency
  extraction method are shown here.}  \centering
\begin{tabular}{c c c c c}
\hline
\textbf{Conditions} & \textbf{Difference (ms)} & \textbf{SE} & \textbf{Z-score} & \textbf{t-test p-value} \\ [0.5ex]
\hline
~ & \multicolumn{4}{c}{Primary latency extraction method} \\
\hline
ND/SD & 40.9 & 1.67 & 24.5 & \textless2.4$\times$10$^{-7}$ \\
ND/AD & 49.8 & 2.07 & 24.1 & \textless2.4$\times$10$^{-7}$ \\
SD/AD & 8.92 & 2.24 & 3.98 & 0.000015 \\ [1ex]
\hline
~ & \multicolumn{4}{c}{Alternative latency extraction method} \\
\hline
ND/SD & 37.3 & 2.00 & 24.5 & \textless1$\times$10$^{-3}$ \\
ND/AD & 49.0 & 2.34 & 20.9 & \textless1$\times$10$^{-3}$ \\
SD/AD & 11.6 & 2.45 & 4.73 & \textless1$\times$10$^{-3}$ \\ [1ex]
\hline
\end{tabular}
\label{table:twosample}
\end{table}

\begin{figure}[htb!]
\centering
\includegraphics[width=0.7\textwidth]{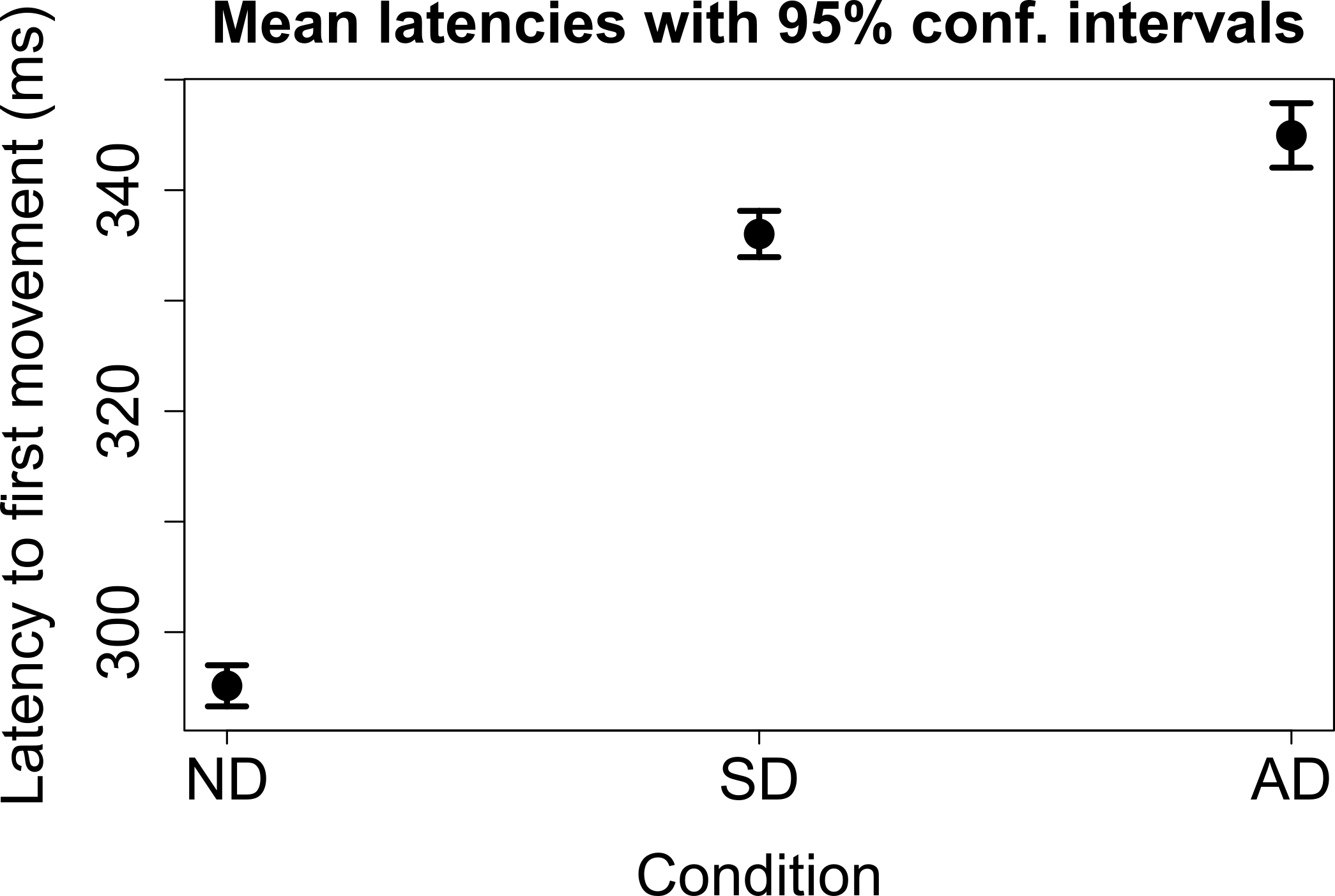}
\caption[Mean latencies] {Mean latencies to first movement in the
  three experimental conditions `No Distractor' (ND), `Synchronous
  Distractor' (SD) and `Asynchronous Distractor' (AD) as determined
  using the primary latency extraction method. In this plot, the
  error bars show 95\% confidence intervals, determined by a bootstrap
  analysis of the sample means.}
\label{meanmad}
\end{figure}

To obtain a better estimate of the probability that the distributions
were drawn from the same population, especially as the apparent effect
size between the SD and AD conditions is small, we applied a
bootstrapped, Studentized t-test. This test rejected the null
hypothesis that distributions were drawn from the same population in
each case: For ND compared with either AD or SD:
p\textless2.5$\times$10$^{-7}$; for AD compared with SD: p=0.000015
(4$\times$10$^{6}$ resamples). The mean latencies, with 95\%
confidence intervals computed from a bootstrap analysis of the sample
means, are shown in Fig~\ref{meanmad}.

% Add figure of mean latency times with error bars here. Using the 95%
% confidence intervals for the error bars.
%% This plot generated in Bootstrap_all.r

%%
%% ND error vs ND no error: p=0.005
%% AD vs AD: p=0.0001
%% SD vs SD: p<0.00002
%%
We analysed the latencies for error movements in each condition to
compare these with latencies for correct movements. In each condition,
the mean latency for error movements was significantly shorter than
for correct movements (p\textless0.00002 SD; p=0.0001 AD; p=0.005 ND) by
about 100~ms. Refer to Table~\ref{table:latencies} for the results.

%% Contains error and non-error. Generated by Bootstrap_two_errornoerror.r
\begin{table}[ht]
\caption{Mean latencies for movements directed towards the target
  (non-error) and for movements which were in error and directed away
  from the target. Standard error and the probability of the two
  samples being drawn from the same population were calculated using
  the bootstrap method described in the text.}  \centering
\begin{tabular}{c c c c c c}
\hline
\textbf{Condition} & \multicolumn{2}{c}{\textbf{Mean latency (ms)}} & \textbf{Difference} & \textbf{std. err.} & \textbf{p value} \\ [0.5ex]
\multicolumn{1}{c}{~} & \textbf{Non-error} & \textbf{Error} & \multicolumn{3}{c}{~} \\
\hline
No Distractor           & 294.9 & 218.9 &  -76  & 21 & p=0.005\\
Synchronous Distractor  & 335.9 & 274.4 &  -61  & 4  & p\textless0.00002 \\
Asynchronous Distractor & 344.9 & 229.0 & -116  & 19 & p=0.0001 \\ [1ex]
\hline
\end{tabular}
\label{table:latencies}
\end{table}

% Regarding response time vs. spatial separation: Bock & Eversheim
% 2000 referred to in Cisek & Kalaska 2010 p 282
We analysed the asynchronous data alone to answer the following
questions: 1) Does the recency of the last distractor event affect the
latency for a target or distractor event movement?  2) Does the
relative location of the distractor affect target latency? Put another
way, if the distractor is in the opposite direction from the target,
do latencies differ from the cases where the distractor lies in the
same direction as the target? In each case, we found no significant
difference or trend.

\begin{figure}[htb!]
\centering
\includegraphics[width=0.7\textwidth]{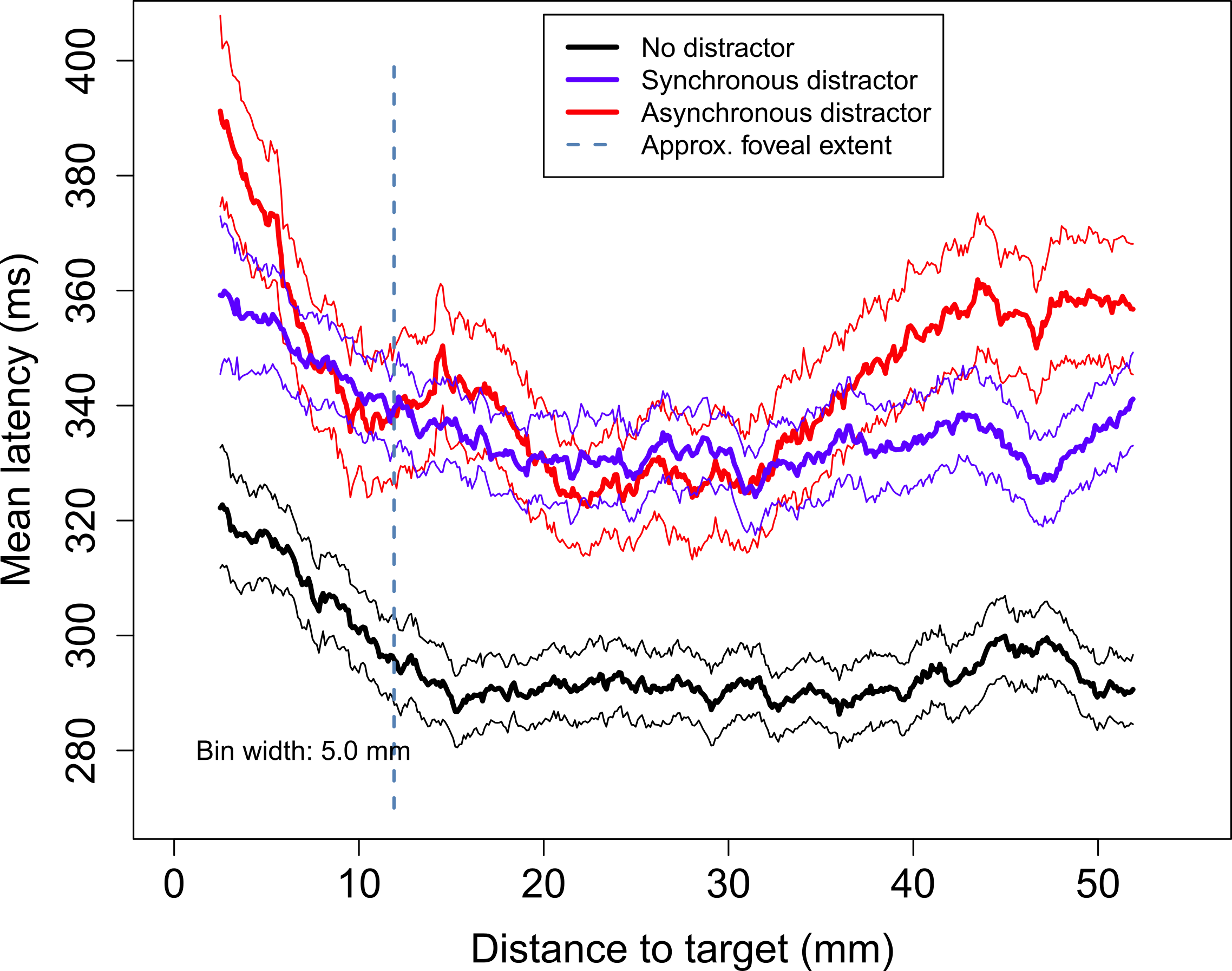}
\caption[Mean latencies vs. distance] {`Moving bin' mean latencies to
  first movement versus distance from stylus to target. A bin of width
  5~mm is moved through the data. At each distance, the mean is
  computed (thicker lines) and the standard error of the mean is
  computed using the bootstrap method. The thinner lines show the
  corresponding 95\% confidence intervals about the mean. The
  approximate location of the foveal region is shown by the dashed
  line, assuming that the eye to screen distance is 400~mm. A marked
  increase in latency is seen for all conditions within the foveal
  region. In the non-foveal region, distance is predictive only for
  the asynchronous distractor.}
\label{latvsdist}
\end{figure}

To examine whether the latency has a dependence on the distance
between the stylus starting position and the target, we arranged all
the latency values in order of distance to target and then computed a
moving mean, using a bin with a width corresponding to 5~mm on the
screen. At each bin position, we compute both the mean, and the 95\%
confidence interval by bootstrapping. Fig~\ref{latvsdist} shows the
result of this analysis. We assume an approximate distance from eye to
tablet of 400~mm, and take the fovea as having an effective radius of
1.7$\deg$, at which point the cone density has dropped to 1/2e of its
maximum~\cite{hirsch_spatial_1989} and rod density is comparable with
cone density~\cite{jonas_count_1992}. This boundary is shown as a
dashed line in Fig~\ref{latvsdist}. The graph shows that the mean no
distractor latency is shorter than the latencies in the distractor
conditions regardless of distance to target. There is a notable upward
trend in latency at short distances, particularly within the
high-acuity foveal region. At longer and very short distances to the
target (greater than 30~mm or less than 7~mm) asynchronous latencies
are longer. Analysing latency data for targets for which distance is
in the non-foveal region, a regression analysis shows that ND and SD
latencies are independent of distance, whereas the latency in the
presence of asynchronous distractors \emph{is} predicted by distance,
F(1,1846)=14.85, p\textless0.001.

\begin{figure}[htb!]
\centering
\includegraphics[width=0.7\textwidth]{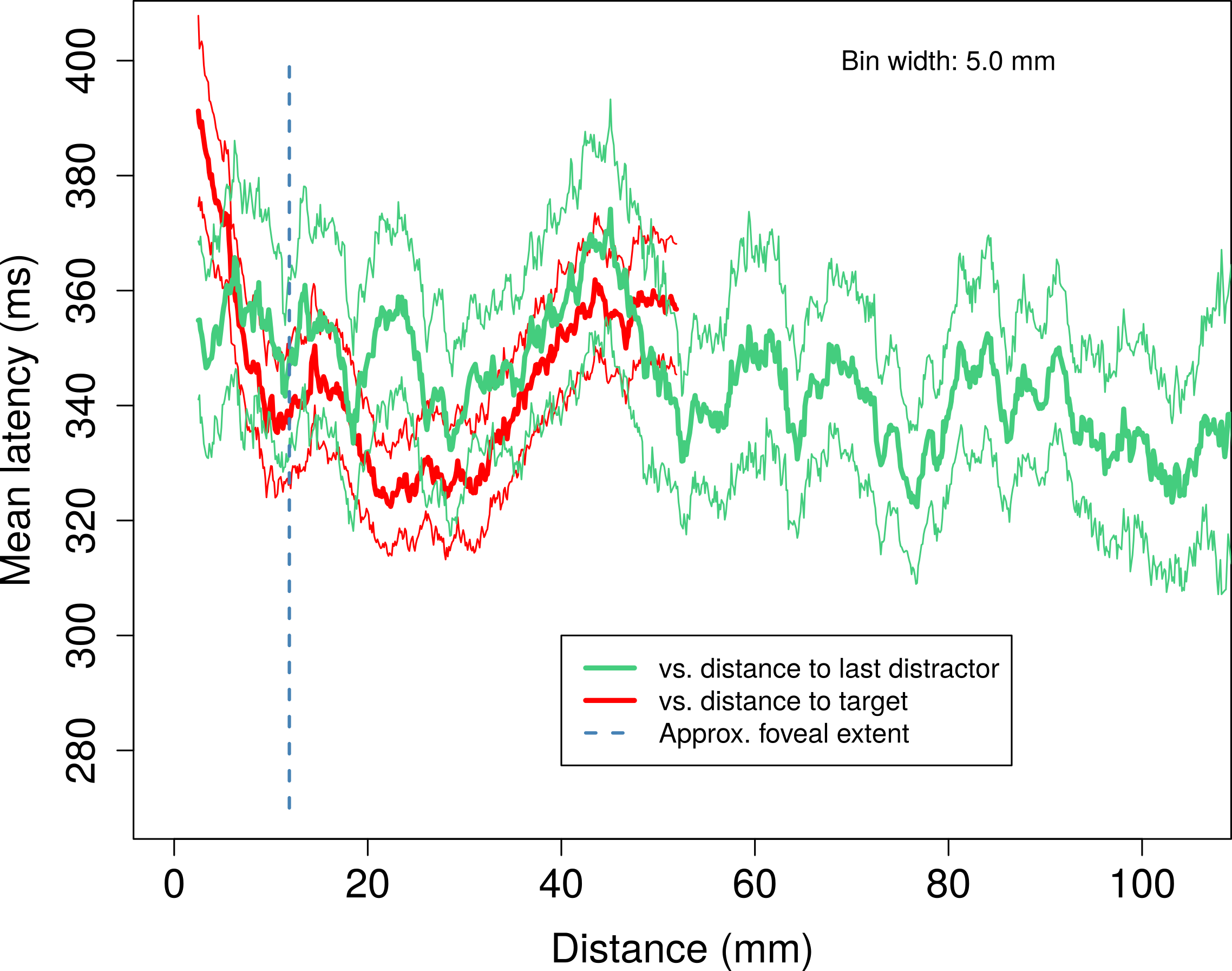}
\caption[Mean latencies vs. distance] {`Moving bin' mean latencies to
  first movement versus distance computed as in Fig~\ref{latvsdist}
  for the asynchronous distractor condition only. Here, latencies to
  first movement are shown with respect to the distance to the target
  (red), which is the same data as the red plot in
  Fig~\ref{latvsdist}, along with latencies to first movement plotted
  with respect to the distance to the distractor which was visible
  when the target event occurred. As in Fig~\ref{latvsdist}, the
  approximate location of the foveal region is shown by the dashed
  line.}
\label{latvslastdist}
\end{figure}

We repeated this analysis to determine if there is any correlation
between the absolute distance to the last distractor in the
asynchronous condition and the latency to first
movement. Fig~\ref{latvslastdist} shows the result of this analysis
and suggests a possible trend that as the last distractor becomes more
distal, the latency reduces very slightly. This trend is not
statistically significant, although a linear model reproduces the
trend for this tiny effect (Slope -0.005 ms/mm, F(1,2230)=2.799,
p=0.094).

\subsection*{Error rates}

We defined the error rate, $R_{C}$, for an individual completing
the line task in condition $C$ as \emph{the number of movement errors
  per target event}:
\begin{equation}\label{eq:error_rate_defn}
R_{C} = \frac{E_{C}}{T}
\end{equation}
$E_{C}$ is the total number of movement errors during the task
and $T$ is the number of target events (excluding those
omitted from the latency extraction).
Note that according to this definition, the asynchronous distractor
error rate, $R_{AD}$, could have a value exceeding 1, as movement
errors could be caused both by target events and by distractor
events. That is, the number of events in the asynchronous condition is:
\begin{equation}\label{eq:num_events_async}
N_{AD} = T + D
\end{equation}
where $D$ is the number of distractor events. The number of events in
the synchronous ($N_{SD}$) and no-distractor ($N_{ND}$) conditions is
smaller, and is given by:
\begin{equation}\label{eq:num_events_sync}
N_{SD} = N_{ND} = T
\end{equation}
Due to these additional distractor events, it was not possible to make
a direct comparison between the error rate for the asynchronous
distractor condition and those for the no-distractor and synchronous
distractor conditions. Whilst the number of events is larger in the
asynchronous condition, the nature of the events is different and so a
definition of the error rate of $R_{C} = {E_C}/{N_C}$ would have
no greater validity than Eq.~\ref{eq:error_rate_defn}.

A Wilcoxon signed
rank test between the ND and SD conditions indicated that the SD error
rate was statistically significantly higher than the ND error rate
(W=23, p\textless10$^{-9}$). Cliff's Delta indicated a large effect
size ($|d|=0.91$).

\begin{figure}[htb!]
\centering
\includegraphics[width=0.7\textwidth]{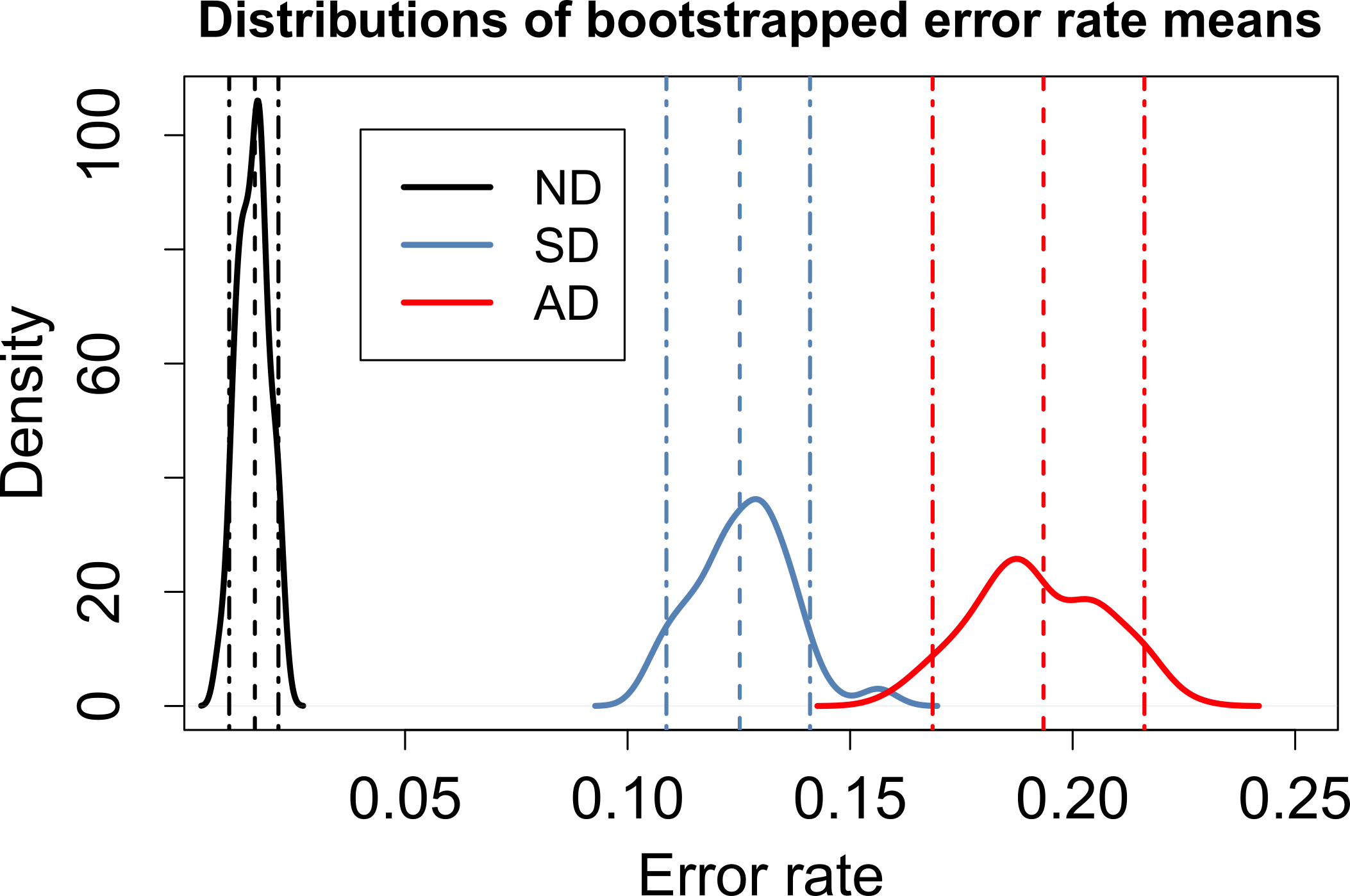}
\caption[Bootstrapped error PDFS] {Probability density functions for
  the error rate sample means computed from 100 data resamples of 55
  data points for each condition. The error rate is computed as the
  proportion of movement errors per target event. Data for the ND
  condition are shown in black and SD in blue. Dashed lines show the
  sample means; dot-dash lines show the 95\% confidence intervals for
  the sample means. It is clear from this graph that
  the synchronous distractor increases the movement error rate
  compared with the no-distractor condition.}
\label{bootstrap_errors}
\end{figure}

The error rates are given in Table~\ref{table:error_rates} and
Fig~\ref{bootstrap_errors}.

% Error rates table
\begin{table}[ht]
\caption{Mean movement error rates with corresponding standard
  deviation and median absolute deviation statistics.}
\centering
\begin{tabular}{c c c c}
\hline
\textbf{Condition} & \textbf{Mean error rate} & \textbf{Std.~dev.} & \textbf{Med.~abs.~dev.} \\ [0.5ex]
\hline
No Distractor & 0.016 & 0.023 & 0.000 \\
Synchronous Distractor & 0.125 & 0.080 & 0.079 \\
Asynchronous Distractor & 0.193 & 0.119 & 0.097 \\ [1ex]
\hline
\end{tabular}
\label{table:error_rates}
\end{table}

\section*{Discussion}

We have presented a new, minimally complex task for the investigation
of target-distractor reaching movements.  The line task constrains arm
movements to a two dimensional plane (because the stylus is
constrained by the surface of the tablet) and guides the subject to
carry out approximately one-dimensional reach movements. Because the
nature of the movements is simple, it is easier to interpret the
behaviour within an action-selection framework in which we consider
the action-space to be one-dimensional.

As a tablet-based experiment, the line task can be conducted either
within or outside the laboratory, with the potential for automated,
large-scale data collection.
Together with the task, we have provided algorithms
which automate the process of extracting latency and movement error
data from the raw trajectories.

The psychophysical results showed that latencies in either distractor
condition were significantly longer than latencies in the
no-distractor condition. This was in broad agreement with previous
studies~\cite{tipper_selective_1992,meegan_visual_1999,pratt_action-centered_1994}.
The results showed that latencies in
the asynchronous distractor condition were slightly, but significantly longer than
those in the synchronous condition and were dependent on the distance
to the target, indicating the value of conducting an easy to deliver
experiment with a large number of participants.

In both of the distractor conditions, latencies for error movements
\emph{towards} the distractor were significantly shorter than correct
movements towards the target. Similarly, in the no distractor
condition, latencies for error movements away from the target were
significantly shorter than correct movements towards the target. Error
rates were significantly higher in the synchronous condition when
compared with the no-distractor condition. The error rate was
significantly higher in the asynchronous condition, although as
discussed, interpretation of this result is difficult because the
event rate in the asynchronous condition is higher than in the other
conditions.
% Have added error rate reporting, as advised by Kevin.

Over the last four decades, much theoretical and experimental work has
been carried out towards a goal of understanding action selection in
the presence of environmental distractors. J.~J.~Gibson introduced the
term \emph{affordance} for an action which is physically possible and
`afforded by the environment'~\cite{gibson_ecological_1979}. In our
task, both target and distractor lines present affordances for
movement. Gibson's work inspired more recent theories of attention and
action selection which we will discuss here; indeed, the theory to
which we will devote most of our discussion is named the
\emph{affordance competition hypothesis}. These more recent theories
make use of neurophysiological data collected from motor and pre-motor
regions of the brain alongside behavioural measurements.

Although we are considering reach movements, it has been natural to
consider action selection and attention in terms of the visual system
and visual perception. Rizzolatti and co-workers presented
psychophysical results both from choice selection and oculomotor
experiments and developed a premotor theory of selective attention. In
their theory, attention is intimitely bound to neural activity which
is preparatory to goal-directed and spatially coded
movements~\cite{rizzolatti_space_1994}. Although the theory makes the
prediction that attended distractors may modify trajectories, it is
not sufficiently quantitative to make predictions about the movement
latencies which are the main topic of this paper.

% Selective attention model
Houghton, Tipper and co-workers produced a large body of work based on
the idea that action and attention are
linked~\cite{tipper_selective_1992,
  houghton_model_1994,houghton_inhibition_1996,tipper_selective_1997,
  tipper_actionbased_1998,tipper_reaching_2001}. In their model of
selective
attention~\cite{houghton_model_1994,houghton_inhibition_1996},
multiple objects are perceived in an `object field'. A competition
occurs to determine the attended object. A `property matching' system
in the model selectively excites target-matching objects and inhibits
non-target objects. The model makes a number of predictions which are
not applicable to the line task (such as the negative priming and
inhibition of return effects~\cite{houghton_model_1994}), but it does
predict that the existence of distractors will slow responses to
targets, in agreement with our results.

% Response vector model
Based on neurophysiological data from Georgopoulus, Kalaska and
co-workers which indicated that neurons in motor
cortex\cite{georgopoulos_neuronal_1986} and parietal
cortex~\cite{kalaska_cortical_1983} produce a signal related to reach
direction, Tipper, Howard and Jackson extended the ideas in the
selective attention model to cover action selection in reach
experiments~\cite{tipper_selective_1997}. In their new model, which is
referred to as the `response vector model' by Welsh and
co-workers~\cite{welsh_movement_2004}, the object field of the
selective attention model becomes a field of potential reach vectors,
interpreted as population codes.  These population codes excite and
inhibit one another via on-centre, off-surround excitation and
inhibition. This assists in the competition between incompatible
actions (such as reach left and reach right). Additional inhibitory
processes selectively inhibit distractors, based on the subject's
goals (e.g.~`ignore the red distractor'). This additional mechanism is
required to prevent intermediate reaches between competing reach
vectors which are too closely aligned for the off-centre inhibition
mechanism to determine a winner.

% Response activation model
Welsh \& Elliott extended a discussion of the response vector model by
considering the timing required by inhibitory processes. The response
vector model does not incorporate a quantitative treatment of the time
required to build up inhibitory suppression and so it is not able to
predict how the effects of synchronous and asynchronous distractor
presentation in the line task might differ and thus cannot help to
explain the small difference in latencies which we detected between
the SD and AD conditions in this work. The `response activation
model'~\cite{welsh_movement_2004} considers that the potential reach
vectors found in overlapping regions of motor cortex participate in a
race. When a movement is expressed, a component of the final movement
will originate from the non-target, predicting trajectory deviations
towards the non-target.  In one experiment in which subjects were
required to reach forwards over distances of up to 0.4~m, Welsh \&
Elliott~\cite{welsh_movement_2004} introduced a `stimulus-onset
asynchrony' between the onset of non-target and target stimuli. They
found that if the non-target (indicated by colour) was illuminated
250~ms or 750~ms before the target, then response times were shortened
by as much as 68~ms compared with a case in which there were no
distractors. They argue that the most likely reason for this response
facilitation is that the non-target induces motor activity which then
aids the target in achieving sufficient activity to produce a
movement. They go on to suggest that if this were true, then the
non-target activity should be seen in trajectory variations. They
presented trajectory data from their experiments which suggest that
motor activity induced by the non-target would be inhibited by 750~ms
after the non-target onset. We did not replicate their result;
instead, we find that our asynchronous distractor condition, which
produces a variable delay between distractor and target (mean(SD)
573(450)~ms), \emph{increases} the latency compared to the no
distractor condition (by 50~ms; see Table~\ref{table:twosample}). Note
that there are important differences between the experiments; Welsh \&
Elliott presented targets over larger distances than those used in the
line task and their subjects had to make reach movements forwards from
their bodies, rather than sideways on a tablet.

A more recent model is the affordance competition
hypothesis~\cite{cisek_cortical_2007}, which is qualitatively similar
to the response vector and activation models but which has closer
links to the neuroanatomy of reach and action selection. Under this
hypothesis, potential actions are continously specified, based on
incoming sensory information, whilst selection processes (both
cortical and sub-cortical) simultaneously determine which are to be
enacted. It extends previous ideas of evidence
integration~\cite{mazurek_role_2003,bogacz_physics_2006} and
sub-cortical action
selection~\cite{redgrave_basal_1999,grillner_mechanisms_2005,houk_action_2007}
by incorporating evidence that activity within motor cortical areas
participates in action selection and specification.

The affordance competition hypothesis contrasts with some more
traditional ideas in cognitive psychology, which describe staged or
serial processes. In the traditional description, sensory information
is first collected to build an internal representation of the external
world.  This representation, along with internal motivations, then
determines or selects the next action. Finally, a process of action
specification is engaged, controlling muscle activation to achieve the
desired action. The difficulty with this framework is that neural
activity relating to sensory information, motor specification and
cognitive value judgements can be found within the same brain nuclei
and are seen to build up concurrently within a
task~\cite{cisek_neural_2005,schall_neural_1993}.

The affordance competition hypothesis suggests that action selection
begins at an early stage within motor cortical brain regions with
activity building up in networks which specify multiple potential
actions. Cortical and subcortical~\cite{humphries_role_2002} processes
take part in supression of non-optimal actions. The model of reaching
decisions presented by
Cisek~\cite{cisek_cortical_2007,cisek_integrated_2006} and supported
by neural activity reported in \cite{cisek_neural_2005} shows activity
in directionally sensitive neurons in posterior parietal cortex (PPC),
dorsal premotor cortex (PMd), primary motor cortex (M1) and in `colour
sensitive' pre-frontal cortex (PFC). A spatial cue is first presented
which indicates two possible direction options. With this cue,
activity is first seen to build in the neurons in PPC and PMd which
encode these two directions. Later, a colour cue indicates which
direction will provide the reward. On presentation of this colour cue,
activity for the correct direction is driven by activity in
`colour-sensitive' pre-frontal cortex, whilst neuronal activity
encoding incorrect directions is supressed. Finally a `go' cue causes
a further increase in activity in PMd and ungates the connection from
PMd to primary motor cortex (M1), providing the motor command. The
model represents a type of sequential evidence accumulation
model~\cite{bogacz_physics_2006}, in that evidence for competing
actions is repeatedly sampled until one action reaches a threshold and
is un-gated.

We now see fit to interpret our task in this framework. In the
synchronous distractor condition of the line task, each event is a
two-alternative forced choice between potential moves towards red and
cyan lines. It is similar to the experimental condition described in
Fig~3(c) in \cite{cisek_cortical_2007} in which two choices are first
presented, then a colour cue indicates the correct target. In our
task, the cyan line is always the target, and so we would expect the
details of the evidence accumulation to differ.  We might expect to
see activity in neurons encoding a move towards the cyan target line
build up faster than activity in those encoding a move towards the red
distractor. The direction of movement made, and whether or not this is
an error will depend on the \emph{rate} at which evidence of the
colour of the lines builds up, the \emph{magnitude of the activity
  threshold} required to un-gate a movement and the \emph{amplitude of
  noise} within the system. The no-distractor condition provides a
control condition in which there is no activity (other than noise) in
competing neurons. The model of Cisek~\cite{cisek_cortical_2007}
predicts that the presence of the distractor will increase the latency
to first movement in the synchronous distractor condition because
cross-inhibition between competing potential actions (`move left' and
`move right') will reduce the rate of evidence accumulation in motor
cortical areas.
% Cisek has FCortex providing the additional inhibition of the
% Houghton and Tipper model.

It is not clear, however, that the affordance competition hypothesis
predicts a shorter latency for error movements; only that evidence for
error movements can be detected as early as activity is seen in PMd
and PPC. The response vector/activation models are similarly unable to
predict this finding. To explain this finding we would require either
that the rate of evidence accumulation caused by the distractor line
is greater than the rate caused by the target line, or that the
threshold to un-gate movement differed for the distractor
line. Neither of these seem plausible, though a colour effect should
be ruled out by reversing the colours used for target and distractor
and repeating the measurements.

It is interesting to consider what Cisek's model would predict for in
the two different distractor conditions and whether it has the ability
to predict the latency difference which we have observed. In Cisek's
model, visual input is passed into PPC, with a short time constant of
0.3~s, and also into colour-sensitive PFC, with a much longer time
constant of 100~s (See Fig~1.A from \cite{cisek_integrated_2006} and
the related supplementary material from that paper). We would expect
to see activity in the
red-sensitive PFC relating to the distractor line and in the blue
sensitive PFC for the target line along with activity for both lines
in PPC. The PFC output interacts multiplicatively with PPC output to
produce input to PMd layers of the model. Given the long time constant
in the PFC, it could be argued that activity from previous distractors
would persist during the presentation of a new target line and slow
the decision making process in PMd. However, the behaviour of PFC is
not modelled in detail in the Cisek model, and instead, the
determination of which colour refers to a target is made by providing
excitatory input to one colour-sensitive PFC population or the
other. In the Cisek modelling, this occurs on the presentation of the
colour code (a central red or blue circle). Given that our experiment
is carried out in blocks, and the target is always blue, there would
be a constant excitatory input to the blue-sensitive PFC, which would
suppress activity in red PFC (red PFC neurons inhibit blue PFC neurons
and vice-versa). It is hard to see how much activity could build up in
PPC relating to the distractor, due to the multiplicative enhancement
of the signal to PMd, and the feedback from PMd to PPC. To apply the
Cisek model to our task, it would likely be necessary to increase the
noise level in the PFC relative to the amount of constant excitatory
input, otherwise the model would not reproduce the errors seen in the
experiment. In any case, we might expect the concurrent presentation
of the distractor in the synchronous condition to have a more
competitive effect, slowing the decision making process, when compared
with the asynchronous condition; a reverse of the observed trend.

The threshold required for activation of M1 is thought to be governed
by activity in related thalamo-cortical loops through the basal
ganglia~\cite{parent_functional_1995,humphries_role_2002}. Channels
have been identified in these loops~\cite{zarzecki_distribution_1991},
and have been postulated to carry motor
commands~\cite{graybiel_basal_1994,doya_complementary_2000,middleton_basal_2000}.
Output nuclei of selected channels through the basal ganglia are
disinhibited, enabling the execution of actions via
brainstem~\cite{roseberry_cell-type-specific_2016} and
cerebellar~\cite{fortier_cerebellar_1989} activity. The same reach
movement could be selected in any of the no-distractor, synchronous
and asynchronous conditions of the line task; for example, the target
could specify a movement to the left of 40~mm. In all conditions, the
same channel in basal ganglia is (presumably) selected. The
architecture of the basal ganglia ensures that the threshold required
to select the channel is kept relatively
constant~\cite{gurney_computational_2001,bogacz_basal_2007},
regardless of the number of competing inputs. Using this argument, we
suggest that an increase in the un-gating threshold activity in PMd
which would be observed in the asynchronous condition of the line task
is unlikely.

In order to explain a lengthened asynchronous latency, the evidence
accumulation seen in PMd should be slowed, which could be achieved
with a blanket inhibition of PMd provided by higher cortical areas.
This could be driven by experiencial learning that in the asynchronous
condition, early evidence from edge detection regions of visual cortex
does not always imply the need for a movement, as it does in the
synchronous condition.

%Fig~\ref{latvsdist} indicates that there is a further nuance to our
%result for the asynchronous distractor condition; that the distance to
%the target has an effect on the mean latency. As the distance to the
%target increases, the mean latency increases.

% Future ideas:
% Could investigate effect of varying the rate of distractor events in
% the AD condition.
%
% Could look at the effect of colour differentiation - possible reason
% for the latency differences seen between the groups?

The simplicity of the design of the line task means that the influence
of distractor interference on action selection mechanisms can be
probed in large-scale subject populations.  Tracking tasks are
valuable for the study of sensory motor function, enabling the
collection of objective and quantitative data, as well as producing
detailed information about task performance. It is not necessary to
provide training to carry out tracking tasks, as making point or reach
movements is entirely natural. Tracking tasks can be equally well
completed by patients with neurological
disorders~\cite{hocherman_recruitment_2004} as by healthy subjects.

The employment of experimental tracking tasks means that considerable
control can be held over stimuli, such as the signalling of targets
and distractors~\cite{watson_two-dimensional_1997}. In spite of this,
Watson and colleagues~\cite{watson_two-dimensional_1997} suggest that
two dimensional tasks provide more precise measurements and
considerably more information regarding sensory-motor function.
Moreover, it is suggested that two dimensional tracking tasks are more
sensitive than one dimensional tracking tasks in their ability to
detect dysfunctional motor performance within Parkinson's
patients. The apparent limitation of using one dimensional
rather than two dimensional pursuit tracking tasks should be
considered in future research.

The line task can be accused of lacking ecological validity, in
that it does not represent tasks common to everyday life: In the line
task, reaches are small (limited by the tablet size) and follow one
dimensional lines as targets; in daily life, reaches may be large (up
to 0.5~m) and are usually made to three dimensional targets for
grasping. The line task purposefully trades ecological validity for a
simple environment with few potentially confounding factors, as
discussed in the introduction.

Future implementations of the line task could vary aspects of the
distractor's parameters. For example, the rate of distractor event
presentation, the number of distractor lines or distractor and target
thicknesses could be varied. A future experiment should be run to
exclude the possibility that there is a colour effect, with the target
and distractor colours being reversed for some subjects, or between
conditions. Gaps between target and distractor presentation could be
investigated, analogous to the `gap, step and overlap' paradigm used
in saccadic eye movement experiments. Investigations in which the
target position was updated \emph{during} the initial stylus movement
could give insight into selection and online error correction within
motor systems. Whilst not discussed here, the dynamics of the
trajectories recorded by the line task could be analysed in detail,
with particular attention being paid to the `first smooth
movement'. In order to more carefully study the effect of the distance
to the target, it would be necessary to record or control the distance
from each subject's eyes to the tablet surface. Consideration should
always be given to learning effects in each condition of the line
task~\cite{song_automatic_2007}. Here, the order in which conditions
were presented was randomised, but it may be preferable to implement
an independent measures design, with each group completing the
opposite order of distractor conditions.

In conclusion, the line task is a tablet-based application that can be
used to record data for healthy or neurologically impaired subjects in
the laboratory, clinic or at home. As such, it is an effective and
convenient tool for the detailed investigation of action selection for
reaching movements.

\section*{Supporting Information}

\subsection*{Supplementary text}

\label{linetask_omitreasons}
{\bf S1 Appendix. A list of omit reasons.} A list of the reasons for which it may
have been impossible to extract a latency and error measurement for an
event.

\label{linetask_anova}
\paragraph{S2 Appendix. Statistical analysis.} PDF version of statistical analysis
ipython notebook.

\section*{Acknowledgments}

This work was partially funded by the EC FP7 project NoTremor ---
Virtual, Physiological and Computational Neuromuscular Models for the
Predictive Treatment of Parkinson's Disease, Grant Agreement
No.~610391.

\nolinenumbers

\end{document}